\documentclass[a4paper,11pt,dvipsnames]{article}

\pdfoutput=1 

\usepackage{jheppub} 

\usepackage{shuffle}
\usepackage{comment}
\usepackage[T1]{fontenc} 
\usepackage{tikz}
\usepackage{tikz-3dplot}
\usepackage{xcolor}
\usetikzlibrary{patterns, decorations.markings, arrows}
\usetikzlibrary{arrows.meta}
\usepackage{mathabx}
\usepackage[export]{adjustbox}
\allowdisplaybreaks
\usepackage[backgroundcolor=green!5, linecolor=blue!60]{todonotes}

\newcommand{\la}{\langle}
\newcommand{\ra}{\rangle}
\newcommand{\ang}[1]{\langle #1\rangle}

\title{Loops of Loops Expansion in the Amplituhedron}

\author{Taro V. Brown,}\emailAdd{tvbrown@ucdavis.edu}
\author{Umut Oktem,}\emailAdd{ucoktem@ucdavis.edu}
\author{Shruti Paranjape,}\emailAdd{sparanjape@ucdavis.edu}
\author{Jaroslav Trnka}\emailAdd{trnka@ucdavis.edu}

\affiliation{Center for Quantum Mathematics and Physics (QMAP),\\Department of Physics, University of California, Davis, CA 95616, USA}

\abstract{We study a novel geometric expansion for scattering amplitudes in the planar sector of ${\cal N}=4$ super Yang-Mills theory, in the context of the Amplituhedron which reproduces the all-loop integrand as a canonical differential form on the positive geometry. In a paper by Arkani-Hamed, Henn and one of the authors, it was shown that this result can be recast in terms of negative geometries with a certain hierarchy of loops (closed cycles) in the space of loop momenta, represented by lines in momentum twistor space. One can then calculate an all-loop order result in the approximation where only tree graphs in the space of all loops are considered. Furthermore, using differential equation methods, it is possible to calculate and resum integrated expressions and obtain strong coupling results. In this paper, we provide a more general framework for the `loops of loops' expansion and outline a powerful method for the determination of differential forms for higher-order geometries. We solve the problem completely for graphs with one internal cycle, but the method can be used more generally for other geometries.}

\begin{document} 

\maketitle

\section{Introduction}

Planar ${\cal N}=4$ super Yang-Mills (SYM) theory is a great playground for testing new theoretical ideas, including new methods for and approaches to the calculation of scattering amplitudes \cite{Arkani-Hamed:2022rwr}. This involves many different studies of the amplitudes - Wilson loop duality and dual conformal symmetry \cite{Alday:2007hr, Drummond:2007aua, Brandhuber:2007yx, Drummond:2007cf}, symbols \cite{Goncharov:2010jf, Golden:2013xva, Golden:2014xqa}, hexagon bootstrap and antipodal duality \cite{Dixon:2021tdw, Dixon:2011nj, Caron-Huot:2019vjl, Caron-Huot:2019bsq, Basso:2020xts, Dixon:2022rse}, flux tube S-matrix approach \cite{Basso:2015uxa, Basso:2014hfa, Basso:2013vsa}, all-loop recursion relations and momentum twistor integrands \cite{Arkani-Hamed:2010pyv,ArkaniHamed:2010kv}, the geometric reformulation of the problem using the positive Grassmannian and on-shell diagrams \cite{Arkani-Hamed:2009ljj, Arkani-Hamed:2009nll, Arkani-Hamed:2009kmp, Arkani-Hamed:2009pfk, Mason:2009qx, Arkani-Hamed:2016byb,Herrmann:2016qea,Bourjaily:2016mnp,Paranjape:2022ymg}, and the Amplituhedron \cite{Arkani-Hamed:2013jha, Arkani-Hamed:2013kca, Arkani-Hamed:2017vfh}, see also \cite{Arkani-Hamed:2021iya,Arkani-Hamed:2018rsk,Franco:2014csa,Dian:2021idl,Dian:2022tpf,Herrmann:2020qlt,Kojima:2020tjf,Rao:2019wyi,YelleshpurSrikant:2019meu,Langer:2019iuo,Heslop:2018nht,Ferro:2016ptt,Ferro:2016zmx,Dennen:2016mdk,Bern:2015ple,Ferro:2015grk,Bai:2015qoa,Arkani-Hamed:2014dca,Bai:2014cna,Prlina:2017azl,Prlina:2017tvx,Salvatori:2018fjp,Dixon:2016apl,Herrmann:2020oud,Arkani-Hamed:2023epq} and \cite{Arkani-Hamed:2017mur,Arkani-Hamed:2019mrd,Arkani-Hamed:2019plo,Arkani-Hamed:2019vag,Arkani-Hamed:2023lbd,Arkani-Hamed:2023bsv,Arkani-Hamed:2023swr} for exciting recent work on geometries in other theories. 

In the Amplituhedron picture, the all-loop integrand is obtained as a canonical differential form on a positive geometry. This translates the physics problem of particle scattering into the mathematical problem of triangulation of a geometric space (or some other method of calculating the canonical form). The Amplituhedron has been studied from a purely mathematical perspective as a natural generalization of the positive Grassmannian with many exciting properties \cite{Lam:2014jda,Karp:2016uax,Karp:2017ouj,Galashin:2018fri,Ferro:2018vpf,Lukowski:2019kqi,Lukowski:2020dpn,Gurdogan:2020tip,Mohammadi:2020plf,Lauda:2020tee,Karp:2021uap,Parisi:2021oql,Williams:2021zph,Moerman:2021cjg,Even-Zohar:2021sec,Blot:2022geq,Even-Zohar:2023del,Akhmedova:2023wcf}. It has been used to obtain a number of interesting results, but the main computational ambition is still left unresolved: calculating all-loop integrands. While this is something very difficult (or impossible) to do using more traditional methods, the Amplituhedron provides a clear mathematical definition of the problem, which still remains unresolved. There is also an interesting redefinition of the Amplituhedron in momentum space, ie. Momentum Amplituhedron, which provides a great window into generalizations of this geometric picture to theories beyond planar limit \cite{Damgaard:2019ztj,Lukowski:2020bya,Ferro:2020lgp,Damgaard:2020eox,Damgaard:2021qbi,Lukowski:2021amu,Lukowski:2021fkf,Lukowski:2023nnf}. Recently, the Amplituhedron pictures has been also obtained for the ABJM theory \cite{He:2023rou,He:2023exb,He:2022cup,He:2021llb,Huang:2021jlh,Henn:2023pkc}.

The simplest case is a four-point scattering amplitude which has a particular simple integrated form -- the leading IR divergence is given by the cusp anomalous dimension \cite{Beisert:2006ez} and the full kinematical part is given by the Bern-Dixon-Smirnov (BDS) ansatz \cite{Bern:2005iz}. Yet the integrand at higher loops is incredibly complex and involves a lot of physical information hidden in its cuts. Explicit calculations have allowed the determination of the answer up to 7-loops using unitarity methods \cite{Bern:2006ew,Bern:2007ct,Bourjaily:2011hi} and up to 10-loops using a certain hidden symmetry of the correlator in the dual Wilson loop picture \cite{Eden:2011we,Eden:2012tu,Bourjaily:2015bpz,Bourjaily:2016evz}. The Amplituhedron definition of the four-point problem is strikingly simple and can be written on two lines: at $L$-loops the integrand is a collection of $L$ lines in momentum twistor space which satisfy certain quadratic conditions. While the problem can be easily stated in geometric language, it is still extremely difficult and a general approach has not been found.

In \cite{Arkani-Hamed:2021iya}, Arkani-Hamed, Henn and one of the authors proposed a curious expansion using \emph{negative geometries} of the $L$-loop integrand of the logarithm of the amplitude (which is essentially the same as the $L$-loop amplitude up to products of lower-loop terms). These geometries are analogous to the positive Amplituhedron geometry up to a sign change which has major implications in the structure of the answer. In particular, when integrated over real Minkowski space they only yield a very mild $1/\epsilon^2$ divergence (in dimensional regularization). This can be seen in the singularity structure of the canonical form (i.e. cut structure of the integrand) and the behavior in the collinear regions. In this framework, the $L$-loop integrand is then a sum of canonical forms for negative geometries. There is an emergent organizational principle among these geometries related to their graph-theoretical representation and the counting of internal loops (or cycles). In \cite{Arkani-Hamed:2021iya}, the authors considered only geometries with no internal cycles, and solved the problem exactly for this case. Moreover, they showed that when integrating over all-but-one loops, the result is IR finite and equal to a  Wilson loop with a Lagrangian insertion \cite{Alday:2011ga,Engelund:2012re,Engelund:2011fg,Alday:2013ip,Henn:2019swt,Chicherin:2022bov,Chicherin:2022zxo}. Using differential equation methods they managed to resum these contributions and get an exact result (in this geometric approximation). 

In this paper, we continue this effort and consider geometries with one internal cycle and solve the problem completely for all canonical forms to any loop order $L$. We show that there is a powerful method for determining the form, without doing any triangulation, which instead relies on imposing constraints from vanishing cuts and cancellation of double poles. We outline how this method can be used for the calculation of geometries with more internal cycles and provide an explicit example. The main open problem is the analogue of the differential operator in \cite{Arkani-Hamed:2021iya} which could be used for these `one-loop' geometries, and to see whether similar resummation procedure can be used. The paper is organized as follows: in section \ref{sec:background}, we provide some background on the Amplituhedron, in section \ref{sec:neggeo} we review the basic ideas behind negative geometries and the solution for all tree graphs of \cite{Arkani-Hamed:2021iya}. In section \ref{sec:conditions}, we enumerate consistency conditions on canonical forms and in section \ref{sec:result} we use these conditions to solve all one-cycle geometries. In section \ref{sec:form}, we provide a useful decomposition for a geometry with more cycles and in section \ref{sec:polylogs} we comment on the integrated IR finite expressions. We conclude with some open problems in section \ref{sec:outlook}.

\section{Background}
\label{sec:background}

The all-loop integrand in planar ${\cal N}=4$ SYM theory can be obtained from a canonical differential form on the \emph{Amplituhedron} space. The original formulation was based on a certain projection from the positive Grassmannian (and its generalization) through the constant positive matrix $Z$ representing external kinematical data \cite{Arkani-Hamed:2013jha}, which was later reformulated in a different more topological language using inequalities and sign flip patterns \cite{Arkani-Hamed:2017vfh}. Apart from the momentum twistor formulation that we use here, there is also a momentum space formulation referred to as the Momentum Amplituhedron \cite{Damgaard:2019ztj}. In this paper, we use the original momentum twistor formulation and focus on four-point amplitudes to all orders. 

\subsection{Definition of the four-point Amplituhedron}

The four-point $L$-loop Amplituhedron \cite{Arkani-Hamed:2017vfh} is defined as a configuration of $L$ lines in the momentum twistor space \cite{Hodges:2009hk},
\begin{center}
	\begin{tabular}{cc}
	 \includegraphics[scale=.65]{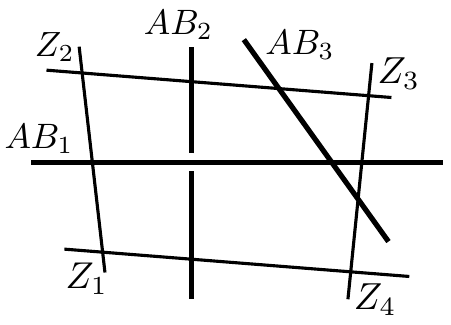}
	\end{tabular}
\end{center}
The external data are represented by four momentum twistors $Z_1$, $Z_2$, $Z_3$, $Z_4$ and must satisfy 
\begin{equation}
\la 1234\ra \equiv \epsilon_{IJKL} Z_1^I Z_2^J Z_3^K Z_4^L > 0\,.
\end{equation}
Each loop is represented by one line $AB_i$ where $A_i$, $B_i$ are two arbitrary points on that line. The configuration space of the line $AB_i$ is restricted to satisfy the following inequalities
\begin{equation}
\la AB_i12\ra, \la AB_i23\ra, \la AB_i34\ra, \la AB_i14\ra>0,\quad \la AB_i13\ra,\la AB_i24\ra<0 \,.\label{oneloopspace}
\end{equation}
We can represent the loop lines using the external momentum twistors, and will often use the standard parametrization
\begin{equation}
\label{eq:param}
    Z_{A_i} = Z_1 + x_i Z_2 - y_i Z_4,\quad Z_{B_i} = Z_3 + z_i Z_2 + w_i Z_4\,.
\end{equation}
In that case, the inequalities reduce to $x_i,y_i,z_i,w_i>0$. Furthermore, for each pair of lines $AB_i$, $AB_j$ we have a mutual positivity condition,
\begin{equation}
\la AB_i\,AB_j \ra > 0\quad \mbox{for each pair $i,j$}\,.
\end{equation}
In the parametrization \eqref{eq:param} this corresponds to 
\begin{equation}
    - (x_i-x_j)(w_i-w_j) - (y_i-y_j)(z_i-z_j) > 0\,.
\end{equation}
These mutual inequalities make the Amplituhedron space very complicated at higher $L$. We define a canonical form $\Omega_L$ with logarithmic singularities on the boundaries of this space, which directly reproduces the $L$-loop integrand form,
\begin{equation}
\Omega_L = {\cal I}_L d\mu_L \quad\mbox{where}\quad d\mu_L = \prod_{i=1}^L \la AB_i\,d^2A_i\ra\la AB_i\,d^2B_i\ra
\end{equation}
where $d\mu_L$ is the integration measure one the configuration of $L$ lines in the momentum twistor space (we often drop this factor for convenience). At the low loop order, we often use a more traditional notation for the loop lines, $AB\equiv AB_1$, $CD\equiv AB_2$, $EF\equiv AB_3$ to make the formulas more transparent. 

\subsection{Calculating the differential form}

The four-point integrand is equal to the canonical differential form with logarithmic singularities on the boundaries of the space. At one-loop the space is extremely simple, given just by a set of inequalities $x,y,z,w>0$ and the dlog form is just
\begin{equation}
\Omega_1 = \frac{dx}{x}\frac{dy}{y}\frac{dz}{z}\frac{dw}{w} = \frac{\la 1234\ra^2d\mu}{\la AB12\ra\la AB23\ra\la AB34\ra\la AB14\ra}\,,
\end{equation}
where $d\mu = \la AB\,d^2A\ra\la AB\,d^2B\ra$ is the integration measure of a line in $\mathbb{P}^3$, analogue of $d^4\ell$ in momentum space. At higher loops, the situation is more complicated and we can pursue multiple strategies to obtain the form. The most straightforward is to triangulate the space into elementary ``simplices'', write the dlog forms for them and sum. 

At two loops, we need to divide the space into eight elementary regions when solving the mutual inequality condition \cite{Arkani-Hamed:2013kca}. The space is defined in the standard parametrization
\begin{equation}
x_1,y_1,w_1,z_1,x_2,y_2,w_2,z_2>0,\quad     - (x_1{-}x_2)(w_1{-}w_2) - (y_1{-}y_2)(z_1{-}z_2) > 0\,.
\end{equation}
We divide the space into eight simplices by giving definite signs to differences $(x_1{-}x_2)$, $(y_1{-}y_2)$ and $(z_1{-}z_2)$. One of them is given by the following inequalities
\begin{equation}
x_1>x_2>0,\,\,y_1>y_2>0,\,\,z_1>z_2>0,\,\,w_1>0,\,\,w_2>w_1 + \frac{(y_1{-}y_2)(z_1{-}z_2)}{(x_1{-}x_2)}\,,
\end{equation}
and the canonical dlog form for this simplex is
\begin{equation}
    \Omega_2^{(1)} = \frac{dx_1}{(x_1-x_2)}\frac{dx_2}{x_2}\frac{dy_1}{(y_1-y_2)}\frac{dy_2}{y_2}\frac{dz_1}{(z_1-z_2)}\frac{dz_2}{z_2}\frac{dw_1}{w_1}\frac{dw_2}{w_2-w_1 - \frac{(y_1-y_2)(z_1-z_2)}{(x_1-x_2)}}\,.
\end{equation}
The poles $(x_1{-}x_2)$, $(y_1{-}y_2)$ and $(z_1{-}z_2)$ are just artifacts of the triangulation, and when summing over all pieces they go away. Summing all the contributions,
\begin{equation}
    \Omega_2 = \sum_{i=1}^8 \Omega_2^{(i)} = \frac{dx_1\,dx_2\,dy_1\,dy_2\,dz_1\,dz_2\,dw_1\,dw_2\,(x_1w_2+x_2w_1+y_1z_2+z_1y_2)}{x_1x_2y_1y_2w_1w_2z_1z_2[(x_1{-}x_2)(w_1{-}w_2)+(y_1{-}y_2)(z_1{-}z_2)]}\,.
\end{equation}
When writing back in terms of momentum twistors we get
\begin{equation}
\Omega_2 = \frac{d\mu_2(\la AB12\ra\la CD34\ra {+} \la AB23\ra \la CD14\ra {+} \la AB34\ra \la CD12\ra {+} \la AB14\ra\la CD23\ra)\la 1234\ra^3}{\la AB12\ra\la AB23\ra\la AB34\ra \la AB14\ra\la ABCD\ra\la CD12\ra\la CD23\ra\la CD34\ra\la CD14\ra}\,, \label{L2}
\end{equation}
where we use the usual notation for two loop lines $AB$ and $CD$, rather than the generic $AB_i$. When expanding the numerator and canceling poles in the denominator we get an expansion of $\Omega_2$ in terms of four double-box integrands ($s$-double box, $t$-double box + symmetrization in $AB$, $CD$). The triangulation strategy becomes difficult already at 3-loops, and a new more sophisticated approach is needed, rather than the naive one above. An alternative strategy is to directly target the numerator $\mathcal{N}_L$ of the form $\Omega_L$. The denominator is known a priori, so we can write the form as 
\begin{equation}
    \Omega_L = \frac{d\mu_L\cdot {\cal N}_L}{\prod_i D_i \cdot \prod_{i,j}\la AB_iAB_j\ra}\,,
\end{equation}
where we denoted the sequence of poles for each loop as 
\begin{equation}
    D_i \equiv \la AB_i12\ra\la AB_i23\ra\la AB_i34\ra\la AB_i14\ra\,.
\end{equation}
Now we can impose various consistency conditions and try to fix ${\cal N}_L$ uniquely. Our strategy is to use \emph{cuts}. There are three kinds of conditions that we need $\mathcal{N}_L$ to satisfy in the end:
\begin{enumerate}
    \item Vanishing on unphysical cuts.
    \vspace{-0.2cm}
    \item Canceling multiple poles generated by the denominator.
     \vspace{-0.2cm}
    \item Reducing to known functions on certain cuts.
\end{enumerate}
In the first case, the denominator generates a singularity on a cut where the amplitude should not have any -- we often call it an \emph{unphysical singularity} -- hence the numerator must vanish. In the second case, the singularity itself might be physical, but the denominator contains a higher-order pole. As the form should only have logarithmic singularities, the numerator must again vanish (sometimes it must even contain a higher degree zero). Both of these conditions are constraints on the \emph{zeroes} of the numerator. The final condition is that the form $\Omega_L$ should reduce to some known function, i.e. the numerator does not vanish on physical cuts, but rather is known and predictable. We use particular ways how to implement these ideas in great detail in later sections.

\subsection{From differential form to amplitude}
\label{sec:logM}
Once we have the integrand form $\Omega_L$ we might want to integrate it to obtain the four-point $L$-loop amplitude,
\begin{equation}
    M_L = \int \Omega_L\,.
\end{equation}
The contour of integration corresponds to real momenta in the momentum space, which translates into a certain complex contour in the momentum twistor space where the lines $AB_i$ live. The $L$-loop amplitude $M_L$ suffers from IR divergences and needs to be regulated. In fact, in dimensional regularization, where we shift the dimension to $4-2\epsilon$, the IR divergence is 
\begin{equation}
 M_L \sim \frac{1}{\epsilon^{2L}}\,.
\end{equation}
It is very well-known that the IR divergence in gauge theories exponentiates, and the invariant new divergence at each order in perturbation theory is at $1/\epsilon^2$ order, the rest is iterative from lower loops. In the planar ${\cal N}=4$ SYM there is an even stronger statement: the Bern-Dixon-Smirnov (BDS) ansatz tells us that the full kinematical dependence of the amplitude (at 4-point and 5-point) is given by an iteration of the one-loop amplitude, and the only new information at every loop order reduces to certain constants \cite{Bern:2005iz},
\begin{equation}
\label{eq:Mwithexp}
M = 1 + \sum_{L=1}^{\infty} g^L M_L(\epsilon) = \exp \left[\sum_{l=1}^\infty g^l \left( f^{(l)}(\epsilon)M^{(1)}(l\epsilon) + C^{(l)}(\epsilon)+E^{(l)}(\epsilon)\right)\right]\,,
\end{equation}
where $g$ is the t'Hooft coupling. The unknown functions $f,C$ and $E$ only depend on $\epsilon$ and some transcendental constants. We can also expand $f^{(l)}=f_0^{(l)}+\epsilon f_1^{(l)} + \epsilon^2 f_2^{(l)}$, where $f^{(l)}_0$ then corresponds to the cusp anomalous dimension $\gamma_{\rm cusp}$. Indeed it is more convenient to work with the logarithm of the amplitude (which is very natural because of the exponentiation in \eqref{eq:Mwithexp}), 
\begin{equation}
\ln M = \sum_{l=1}^\infty g^l f^{(l)}(\epsilon) M^{(1)} (l\epsilon) + C^{(l)}(\epsilon) + E^{(l)}(\epsilon) = \frac{\gamma_{\rm cusp}}{\epsilon^2} + {\cal O}\left(\frac{1}{\epsilon}\right)\,.
\end{equation}
The leading term has a very mild $1/\epsilon^2$ divergence to any loop order and the kinematic dependence is only in the subleading terms. The cusp anomalous dimension $\gamma_{\rm cusp}$ for the planar ${\cal N}=4$ SYM is predicted by integrability \cite{Beisert:2006ez} to all loop orders but a derivation from all-loop amplitudes is still missing. 

It is striking that some hints of the exponentiation of IR divergencies can be seen from the structure of the amplitude integrand, i.e. the $L$-loop canonical differential form $\Omega_L$ on the Amplituhedron. To see that, we explicitly write the perturbative expansion of the logarithm of the amplitude,
\begin{equation}
    \ln M = \ln (1+g^2 M_1 + g^4 M_2 + g^6 M_3 + \dots) = g^2 \ln M_1 + g^4 \ln M_2 + g^6 \ln M_3 + \dots\,,
\end{equation}
where $\ln M_L$ denotes the $L$-loop logarithm corresponding to sums of products of amplitudes. In particular, the first few loop orders are
\begin{align}
\ln M_1 = M_1,\quad \ln M_2 = M_2 - \frac12 M_1^2,\qquad \ln M_3 = M_3 - M_2M_1 +\frac13 M_1^3, \quad \dots \label{log2}
\end{align}
Each term in $\ln M_L$ is maximally $1/\epsilon^{2L}$ IR divergent while in the sum all terms up to $1/\epsilon^2$ cancel. Note that $\ln M_L$ is a combination of products of planar amplitudes, but it itself is not planar, i.e. cannot be expanded as a sum of planar Feynman diagrams. This can be seen already at two-loops where the square of the one-loop amplitude does not correspond to a planar two-loop diagram. However, we can still use global dual variables and momentum twistors and define
\begin{equation}
    \ln M_L = \int \widetilde{\Omega}_L
\end{equation}
where $\widetilde{\Omega}_L$ is the integrand for the $L$-loop logarithm. It can be obtained from summing over integrand forms for amplitudes (and their products), for example,
\begin{align}
    \widetilde{\Omega}_2 &= \Omega_2(AB,CD) - \Omega_1(AB)\Omega_1(CD)\,,\\
    \widetilde{\Omega}_3 &= \Omega_3(AB,CD,EF) - \Omega_2(AB,CD)\Omega_1(EF)- \Omega_2(AB,EF)\Omega_1(CD)\nonumber \\ & \hspace{1cm} - \Omega_2(CD,EF)\Omega_1(AB) + 2 \Omega_1(AB)\Omega_1(CD)\Omega_1(EF)\,,
\end{align}
where the difference in numerical factors here and in (\ref{log2}) is given by symmetrization in the integrand which is completely symmetric in all loop lines $AB_i$. Here we labeled $AB$, $CD$ and $EF$ as three loops, but in general we use the $AB_i$ labeling. The $L$-loop logarithm integrand form $\widetilde{\Omega}_L$ has very special properties in the collinear regions which control IR divergencies \cite{Arkani-Hamed:2013kca}. The idea is that the integrand for the logarithm $\widetilde{\Omega}_L$ is also a canonical form on a certain positive geometry \cite{Arkani-Hamed:2017tmz}. The answer was given in \cite{Arkani-Hamed:2021iya} and indeed, $\widetilde{\Omega}_L$ is a sum of canonical forms on connected \emph{negative geometries}, we will discuss it in section \ref{sec:neggeo}. 

\subsection{IR finite object}

The logarithm of the amplitude $\ln M_L$ is almost an IR finite object, and the IR divergencies only appear if we integrate over all loop momenta -- this is also evident from the cuts of the integrand $\widetilde{\Omega}_L$ which do not allow the loops $AB_i$ to access the collinear region unless all loops are involved in the cut (none of the loops can be left uncut). Hence, we decide to do the maximal ``safe'' integration of $\widetilde{\Omega}_L$ and integrate over $AB_2,{\dots},AB_L$ while freezing (not integrating over) the loop momentum $AB\equiv AB_1$ in the momentum twistor language \cite{Arkani-Hamed:2021iya}. The dual conformal symmetry dictates that the result can only depend on a single cross-ratio for $AB\equiv AB_1$,
\begin{equation}
    z = \frac{\la AB12\ra\la AB23\ra}{\la AB23\ra\la AB14\ra}\,.
\end{equation}
We can then define an IR function ${\cal F}_L(z)$,
\begin{equation}
{\cal F}_L(z) = \int \widetilde{\Omega}_L\,\quad \mbox{where $AB$ is not integrated over}\,.
\end{equation}
Further, we can perform a resummation and introduce a formal non-perturbative function ${\cal F}(g,z)$ by dressing ${\cal F}_L(z)$ with $g^{2L}$ and summing over all $L$,
\begin{equation}
    {\cal F}(g,z) = \sum_{L=1}^\infty g^{2L}{\cal F}_L(z)\,.
\end{equation}
This object has been studied before \cite{Alday:2011ga,Engelund:2012re,Engelund:2011fg,Alday:2013ip,Henn:2019swt,Chicherin:2022bov,Chicherin:2022zxo} and it is equal to the ratio of a null-polygonal Wilson loop with a Lagrangian insertion divided by (an empty) null-polygonal Wilson loop,
\begin{center}
	\begin{tabular}{cc}
	 \includegraphics[scale=.68]{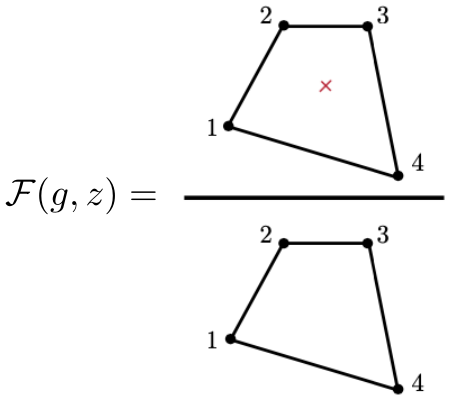}
	\end{tabular}
\end{center}
The object ${\cal F}_L(z)$ has a very natural expansion in terms of \emph{marked} positive geometries as shown in \cite{Arkani-Hamed:2021iya} and we discuss it more in section \ref{sec:polylogs}.

\section{General positive and negative geometries}
\label{sec:neggeo}

We start our discussion with the generalization of the Amplituhedron geometry to a larger class of positive geometries. They correspond to configurations of $L$ lines $AB_i$, where each line satisfies the one-loop Amplituhedron conditions (\ref{oneloopspace}) and in addition, we have a collection of mutual positivity conditions $\la AB_i AB_j\ra>0$ but not necessarily for all pairs of $AB_i$ and $AB_j$. We can graphically represent such geometries as graphs with $L$ nodes -- one for each loop line $AB_i$, where some of the nodes are connected by a blue dashed link, representing a mutual positivity condition $\la AB_iAB_j\ra>0$. Here are some examples of these positive geometries:
\begin{center}
	\begin{tabular}{cc}
	 \includegraphics[scale=.77]{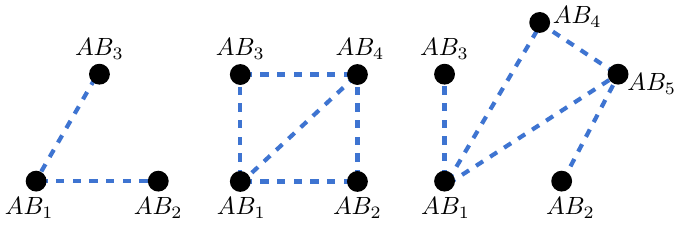}
	\end{tabular}
\end{center}
Sometimes it is convenient to explicitly label $AB_i$ associated with a given node. The most complicated, complete graph, corresponds to the $L$-loop Amplituhedron where the mutual positivity conditions are imposed between any pair of lines. Here are examples of Amplituhedron geometries up to $L=5$:
\begin{center}
	\begin{tabular}{cc}
	 \includegraphics[scale=.8]{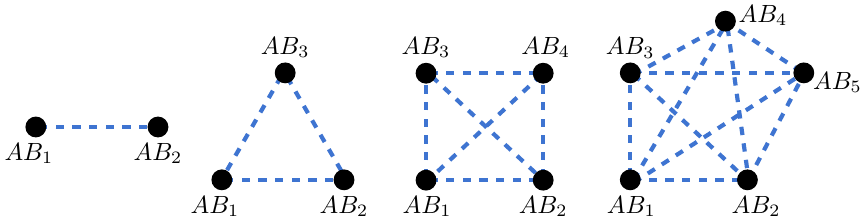}
	\end{tabular}
\end{center}
We could use the graphic notation for specifying a positive geometry, but for us it is more convenient to associate a logarithmic form $\Omega_\Gamma$ on the positive geometry with the graph (rather than just the geometry). We now generalize our discussion slightly and also consider graphs where some of the links correspond to mutual positive conditions $\la AB_i AB_j\ra>0$ and others to mutual negative conditions $\la AB_i AB_j\ra<0$. 
\begin{center}
	\begin{tabular}{cc}
	 \includegraphics[scale=.85]{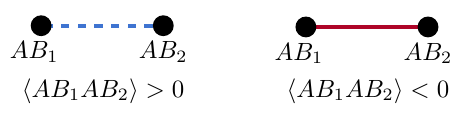}
	\end{tabular}
\end{center}
For nodes that are not connected by any link, the sign of the corresponding term $\la AB_i AB_j\ra$ is unspecified. Here are a few examples of these more general geometries:
\begin{center}
	\begin{tabular}{cc}
	 \includegraphics[scale=.85]{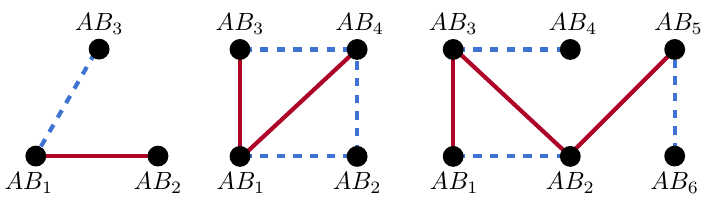}
	\end{tabular}
\end{center}
Note that if the graph is not connected, the differential form is given by a (wedge) product of differential forms for the subgraphs which are connected. For example,
\begin{center}
	\begin{tabular}{cc}
	 \includegraphics[scale=.87]{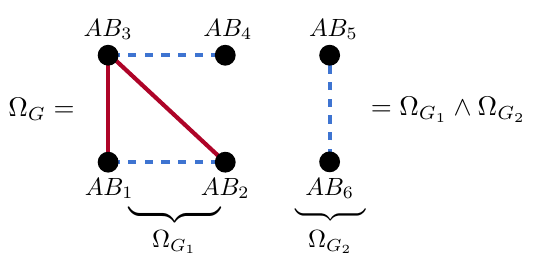}
	\end{tabular}
\end{center}
For simplicity, we skip the wedge symbol $\wedge$ in the following discussion. Note that the graphical notation we introduced is for the differential form $\Omega_G$ of the corresponding geometry. Nonetheless the graph also gives us the geometry itself. To summarize:
\begin{itemize}
\item Each dot is one loop line $AB_i$ which satisfies one-loop Amplituhedron conditions \eqref{oneloopspace}.
\item The dashed blue link between dots $AB_i$ and $AB_j$ imposes a mutual positive condition $\la AB_i AB_j\ra>0$, while the red solid line represents a condition $\la AB_i AB_j\ra<0$. 
\end{itemize}
In some cases, we do not label the $AB_i$ vertices at all -- in this case we implicitly \emph{symmetrize} over all $AB_i$. 

\subsection{From positive to negative}

The positive and negative links are not unrelated, they do satisfy a simple sum rule,

\vspace{-0.3cm}

\begin{equation}
\label{eq:linkrel}
	\begin{tabular}{cc}
	 \includegraphics[scale=.86]{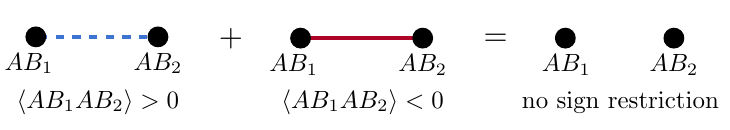}
	\end{tabular}
\end{equation}

\vspace{-0.3cm}

This relation makes sense as a union of two geometries, and as the sum of their canonical forms $\Omega_{G_1}$ and $\Omega_{G_2}$. The resulting geometry is simpler, and we expect the canonical form to have simpler structure as well. This relation makes sense for $L=2$ but can be also used inside an arbitrary complicated graph. For example,
\begin{equation}
	 \includegraphics[scale=.8]{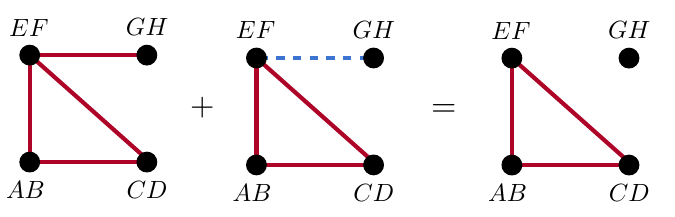}
	 \label{Fig4}
\end{equation}
We can use the relation between the links (\ref{eq:linkrel}), and express all positive links in the graph for the $L$-loop Amplituhedron in terms of negative links and no links. As a result, we express the $L$-loop Amplituhedron form $\Omega_L$ as a sum of canonical forms for \emph{negative geometries} which have only negative links,
\begin{equation}
	 \includegraphics[scale=.7]{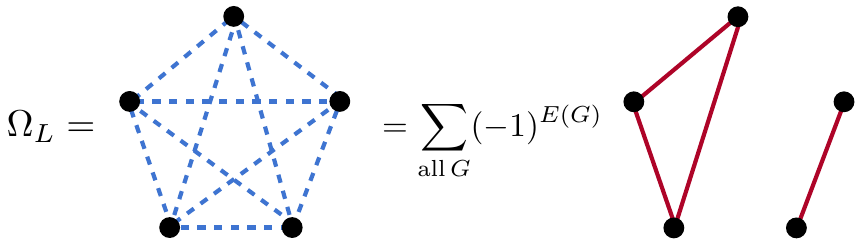}
	 \label{fig4}
\end{equation}
The sum is over all graphs $G$, and $E(G)$ is the number of edges in the graph $G$. This particular relation will be important later on in our discussion. Note that the relation (\ref{fig4}) is not between geometries (the geometries on both sides of the equation are very different, with different positive/negative links) but between their canonical differential forms. The right hand side of (\ref{fig4}) contains all graphs, connected and disconnected, with any number of links, and also with all possible assignments of loop labels $AB_i$.

\subsection{Logarithm of the amplitude from negative geometries}

Let us now consider a restriction of the sum on the right-hand-side of (\ref{fig4}) and sum only over the differential forms of  \emph{connected} graphs with negative links, and define the form $\widetilde{\Omega}_L$,
\begin{equation}
	\begin{tabular}{cc}
	 \includegraphics[scale=.68]{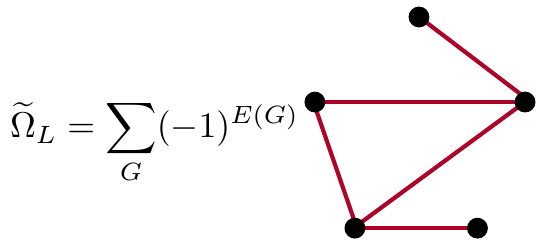}
	\end{tabular}
\end{equation}
where $E(G)$ is the number of edges in the graph. It was shown in \cite{Arkani-Hamed:2021iya} that this object is in fact the integrand for the \emph{logarithm of the amplitude} which we also discussed in section \ref{sec:logM}. This is a simple consequence of the relation between the collection of all graphs and all connected graphs.
\begin{equation}
	\begin{tabular}{cc}
	 \includegraphics[scale=.65]{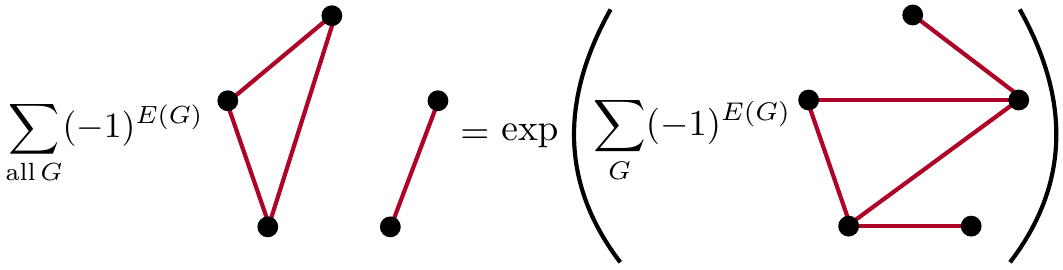}
	\end{tabular}
\end{equation}
More heuristically, any given graph in (\ref{fig4}) can be obtained as the product (at the level of differential forms) of connected graphs, hence knowing the forms for all connected graphs clearly has all information about the forms for all graphs. 

\subsection{Loops of loops expansion}

There is a natural hierarchy in the expansion for $\widetilde{\Omega}_L$ which we presented at the end of the last section: the number of \emph{internal loops} or \emph{cycles}. We will mostly use the word ``cycle'' wherever we can, in order not to overuse the word \emph{loop}, which can be confused with loop lines $AB_i$ (that are represented by nodes in the graphs),
\begin{equation}
\widetilde{\Omega}_L = \widetilde{\Omega}_L^{\rm tree} + \widetilde{\Omega}_L^{\rm 1-cycle} + \widetilde{\Omega}_L^{\rm 2-cycle} + \widetilde{\Omega}_L^{\rm 3-cycle} + \dots  = \sum_{\ell=0}^{\rm max} \widetilde{\Omega}_L^{\ell-{\rm cycle} } \label{loops}
\end{equation}
where $ \widetilde{\Omega}_L^{\ell-{\rm cycle} }$ is the sum of differential forms for all graphs with $\ell$ internal cycles. Tree graphs refer to graphs with no internal cycles. In general for the $L$-loop case with $L$ nodes, the number of links in the graph is $L+\ell-1$. As the maximal number of links is $L(L{-}1)/2$, the maximal number of internal cycles is $(L{-}1)(L{-}2)/2$. We call the expansion (\ref{loops}) the \emph{loops of loops expansion} and it is crucial for our discussion. The idea is that for fixed $L$, the expansion in terms of cycles (internal loops) is meaningful. The higher the number of cycles, the more complicated the geometry is and the more complicated the corresponding differential form $\Omega_\Gamma$ is. This is very natural as the number of cycles grows with the number of links in the graph, and clearly the more the links, the more complicated a differential form we expect.

Let us provide some explicit examples. For $L=2$ we only get one tree graph (two nodes and one link), while for $L=3$ we get both a tree graph and a one-cycle graph,
\begin{equation}
\label{Omega3}
	\begin{tabular}{cc}
	 \includegraphics[scale=.8]{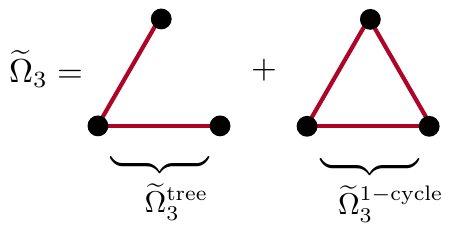}
	\end{tabular} 
\end{equation}
For $L=4$ we get graphs up to 3 cycles. 
\begin{equation}
\label{four2}
	\begin{tabular}{cc}
	 \includegraphics[scale=.8]{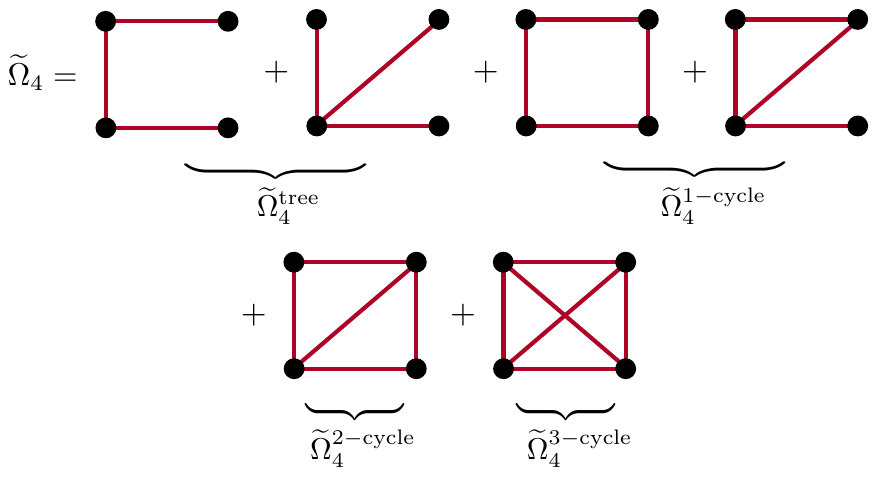}
	\end{tabular}
\end{equation}
where in both (\ref{Omega3}) and (\ref{four2}) we need to label $AB_i$ in all possible ways and sum. The task is to find differential forms for these geometries. The problem of tree forms was solved in \cite{Arkani-Hamed:2021iya}, in this paper we solve for all canonical forms of geometries with one cycle and also outline how to continue to geometries with more cycles. 

\subsection{All tree canonical forms}

The solution for canonical forms for all tree graphs is particularly simple. For $L=2$ we have only one graph topology (two nodes and one link),
\begin{center}
	\begin{tabular}{cc}
	 \includegraphics[scale=.77]{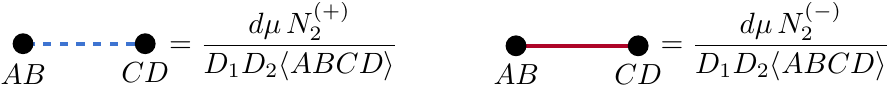}
	\end{tabular}
\end{center}
where we denoted (up to an overall factor $\la 1234\ra^3$ which we drop),
\begin{align}
&D_1 = \la AB12\ra\la AB23\ra\la AB34\ra\la AB14\ra,\quad D_2 = \la CD12\ra\la CD23\ra\la CD34\ra\la CD14\ra\,,\\
&N_2^{(+)} = \la AB12\ra\la CD34\ra + \la AB23\ra\la CD14\ra + \la AB34\ra\la CD12\ra + \la AB14\ra\la CD23\ra\,,\\
&N_2^{(-)} = \la AB13\ra\la CD24\ra + \la AB24\ra\la CD13\ra\,.
\end{align}
The union of both spaces are just two one-loop Amplituhedra, for each $AB$ and $CD$. At the level of the numerators this gives
\begin{equation}
N_2^{(+)} + N_2^{(-)}  = \la ABCD\ra\la1234\ra\,.
\end{equation}
These two numerators will be used as the building blocks for the general $L$-loop case. We define two objects objects $N_{ij}^{(+)}$ and $N_{ij}^{(-)}$ which are two-loop numerators for positive, (resp. negative) links between $AB_i$ and $AB_j$,
\begin{align}
    N_{ij}^{(+)} & \equiv \la AB_i12\ra\la AB_j34\ra + \la AB_i23\ra\la AB_j14\ra + \la AB_i34\ra\la AB_j12\ra + \la AB_i14\ra\la AB_j23\ra\,, \nonumber\\
    N_{ij}^{(-)} & \equiv \la AB_i13\ra\la AB_j24\ra + \la AB_i24\ra\la AB_j13\ra\,.
\end{align}
Then for an arbitrary tree graph with arbitrary assignments of positive and negative links the differential form can be written as 
\begin{center}
	\begin{tabular}{cc}
	 \includegraphics[scale=.75]{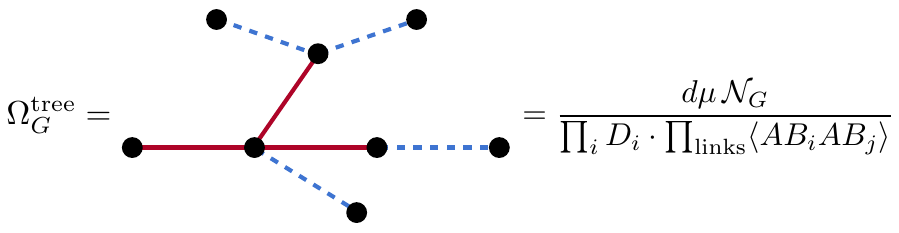}
	\end{tabular}
\end{center}
where the denominator has the expected structure of the product of one-loop numerators
\begin{equation}
D_i = \la AB_i 12\ra\la AB_i 23\ra\la AB_i 34\ra\la AB_i14\ra\,,
\end{equation}
and the mutual propagators $\la AB_i AB_j\ra$ for all links. The numerator carries information about the sign of the mutual conditions, and takes a very simple factorized form:
\begin{equation}
   \boxed{ {\cal N}_G = \langle 1234\ra^{L{+}1} \times \prod_{\rm pos\,links } N_{ij}^{(+)} \times \prod_{\rm neg\, links } N_{ij}^{(-)} }\label{treenum}
\end{equation}
It is worth noting that because of the factorized form of the numerator, any two tree graphs which differ only by a sign of a single link, satisfy the same summation rule as the $L=2$ case (\ref{eq:linkrel}), 
\begin{center}
	\begin{tabular}{cc}
	 \includegraphics[scale=.75]{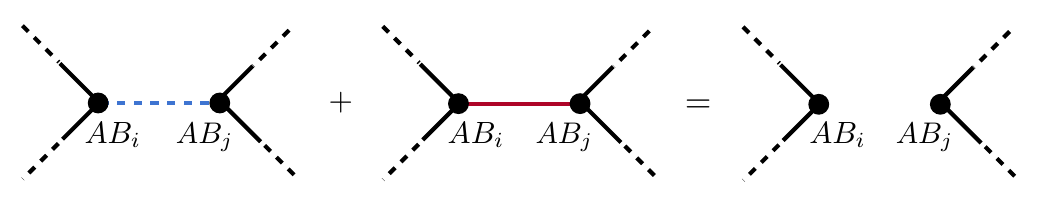}
	\end{tabular}
\end{center}
This is just a trivial consequence of the relation between numerators,
\begin{equation}
N_{ij}^{(+)} + N_{ij}^{(-)}  = \la1234\ra\la AB_iAB_j\ra\,.
\end{equation}
Now for the special case of tree graphs with only negative links the differential form is 
\begin{center}
	\begin{tabular}{cc}
	 \includegraphics[scale=.73]{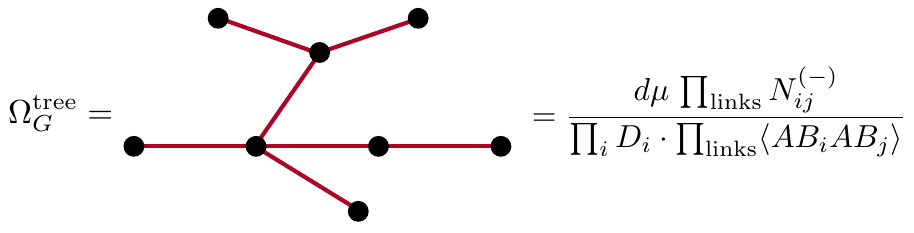}
	\end{tabular}
\end{center}
as also presented in \cite{Arkani-Hamed:2021iya}. Using this formula for an arbitrary tree graph, we can completely determine the first term in the expansion 
\begin{equation}
\widetilde{\Omega}^{\rm tree}_L = \sum_{{\rm trees\,}G} \widetilde{\Omega}_{G}\label{tree}\,.
\end{equation}
Furthermore, in section \ref{sec:polylogs} we will show that we can take the next step and integrate (\ref{tree}) using a differential equation method.

\section{Consistency conditions}
\label{sec:conditions}

Let us now consider a general connected graph,
\begin{center}
	\begin{tabular}{cc}
	 \includegraphics[scale=.77]{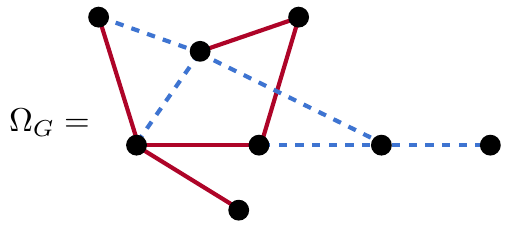}
	\end{tabular}
\end{center}
with some labeling of vertices $AB_i$. There are important consistency conditions on the associated canonical form that come from cuts. Let us remind ourselves that the geometry satisfies 
\begin{equation}
\la AB_i12\ra, \la AB_i23\ra, \la AB_i34\ra, \la AB_i14\ra>0, \quad \la AB_i13\ra,\la AB_i24\ra<0\,,
\end{equation}
for any $AB_i$, and $\la AB_i AB_j\ra\lessgtr 0$ for any link. In this section, we perform cuts on $\Omega_G$, localize one or more loop lines and read off the inequalities of the reduced geometry. 

\subsection{Singlet conditions}

Let us first discuss cuts where only one of the loops gets localized on a cut -- we call them \emph{singlet} conditions. We start with localizing $AB_i$ on the leading singularity
\begin{equation}
\la AB_i12\ra = \la AB_i23\ra = \la AB_i34\ra = \la AB_i14\ra = 0\,,
\end{equation}
which fixes $AB_i =13$ (or $AB_i=24$, the discussion is analogous). If the vertex $AB_i$ is connected to any other vertex $AB_j$ through a positive link, we get
\begin{equation}
\la AB_i AB_j \ra = \la AB_j 13\ra > 0 \quad \mbox{manifestly negative}\,.
\end{equation}
This is in contradiction with the one-loop Amplituhedron conditions for $AB_j$ which require $\la AB_j 13\ra <0$. Hence if $AB_i$ is connected to any other vertex through at least one positive link, the canonical form $\Omega_G$ and the numerator ${\cal N}_G$ must vanish. 
\begin{center}
	\begin{tabular}{cc}
	 \includegraphics[scale=.77]{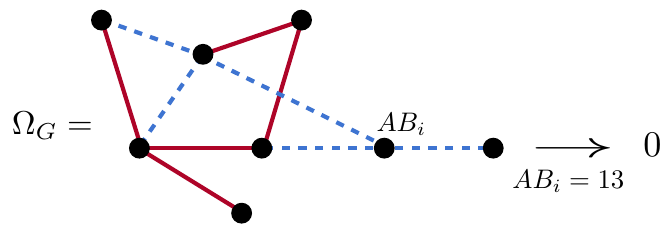}
	\end{tabular}
\end{center}
If the $AB_i$ is connected to other vertices only through negative links, the line $AB_i$ is eliminated from the graph without any additional condition added,
\begin{center}
	\begin{tabular}{cc}
	 \includegraphics[scale=.77]{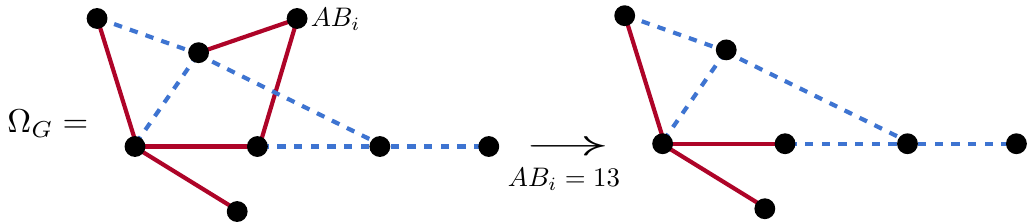}
	\end{tabular}
\end{center}
We refer to this type of condition as {\bf P1} because it is illegal for positive links and involves only one of the loops. Another singlet condition is a composite triple cut 
\begin{equation}
\la AB_i12\ra = \la AB_i23\ra = 0 \,\,\mbox{including the Jacobian}\,,
\end{equation}
which localizes $AB_i = 12 + \alpha 23$, in other words the line $AB_i$ passes through vertex $2$ and lies in a plane $(123)$. In the momentum space this corresponds to a collinear region when $\ell \sim p_2$. The one-loop Amplituhedron condition fixes the sign of $\alpha$,
\begin{equation}
\la AB_i 14 \ra =\alpha \la 1234\ra  > 0 \,\,\xrightarrow \,\, \alpha>0\,.
\end{equation}
Then for a link $\la AB_i AB_j\ra$ we get
\begin{equation}
\la AB_i AB_j\ra = \la AB_j 12\ra + \alpha \la AB_j 23\ra > 0 \qquad \mbox{manifestly positive}\,.
\end{equation}
If $AB_i$ is connected to at least one negative link,
\begin{center}
	\begin{tabular}{cc}
	 \includegraphics[scale=.73]{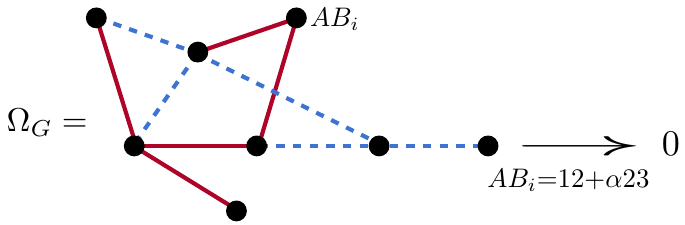}
	\end{tabular}
\end{center}
If all links are positive the residue is 
\begin{center}
	\begin{tabular}{cc}
	 \includegraphics[scale=.73]{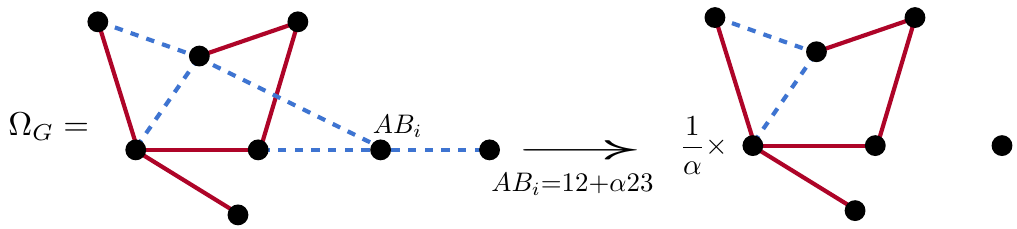}
	\end{tabular}
\end{center}
The $1/\alpha$ factor comes from the Jacobian $1/\la AB_i 14\ra$ which is uncut. We refer to this type of condition as {\bf N1} because it is illegal for negative links and involves only one loop. If the node $AB_i$ is connected through both positive and negative links, then both {\bf P1} and {\bf N1} are illegal and the canonical form should vanish on these cuts.

\subsection{Doublet conditions}

There are also \emph{doublet} conditions which involve cutting two lines, say $(AB)$ and $(CD)$ and finding a contradiction between the fixed sign of $\la ABCD\ra$ and the solution of the cut. In other words, if we have a positive link $\la ABCD\ra>0$ but the cut forces $\la ABCD\ra<0$ this cut is illegal, and vice versa for a negative link $\la ABCD\ra<0$ -- any cut which forces $\la ABCD\ra>0$ is illegal. Examples of the first type of cuts, which force $\la ABCD\ra<0$, are:
\begin{itemize}
\item[\bf (P2a)] The cut when $AB$ cuts lines $12$ and $34$ while $CD$ cuts lines $23$ and $14$. 

\vspace{-0.15cm}

\item[\bf (P2b)] The cut when $AB$ lies in the plane $(123)$ and $CD$ passes through point $2$. This can be achieved by cutting $\la AB12\ra=\la AB23\ra=\la CD12\ra=\la CD23\ra=0$, this is the same as the cut used to detect double poles, but now choosing other cut solution.
\end{itemize}
which can also be easily seen in any parametrization and where we have used the same naming convention as before. We can see from figures
\begin{center}
	\begin{tabular}{cc}
	 \includegraphics[scale=.6]{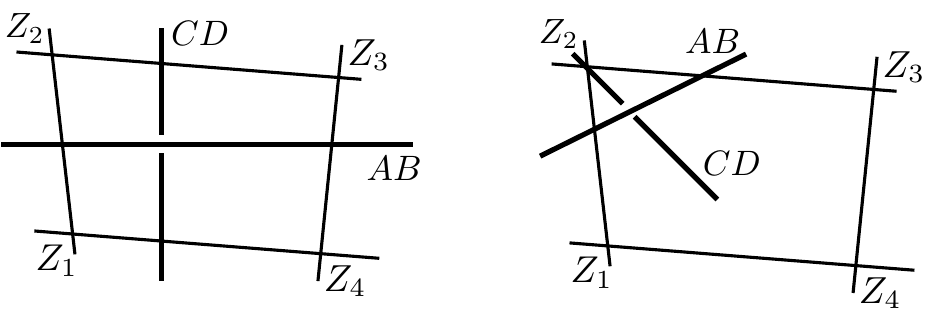}
	\end{tabular}
\end{center}
that both these cuts violate ``two-loop planarity'' and they are not present in any planar diagrams (the loop lines `cross each other'). This makes sense because the $L=2$ space with $\la ABCD\ra>0$ corresponds to a planar two-loop amplitude. There are also the type of cuts which are incompatible with a negative link, ie. force $\la ABCD\ra>0$,
\begin{itemize}
\item[\bf (N2a)] The cut when $AB$ passes through $1$ and $CD$ passes through $3$. To arrive at this configuration we have to cut $\la AB14\ra=\la AB12\ra=\la CD23\ra=\la CD34\ra=0$ and choose an appropriate solution.

\vspace{-0.15cm}

\item[\bf (N2b)] Cutting the same propagators but choosing a different solution when $AB$ is in the plane $(412)$ and $CD$ is in the plane $(234)$ we again arrive at the illegal configuration.
\end{itemize}
\begin{center}
	\begin{tabular}{cc}
	 \includegraphics[scale=.62]{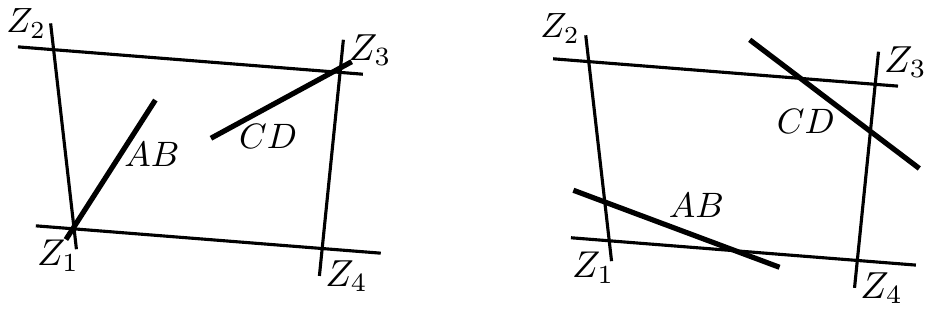}
	\end{tabular}
\end{center}
Here we have less intuition of what this means from a physics perspective, but they are nevertheless illegal for the $\la ABCD\ra<0$ space, which for $L=2$ is $\widetilde{\Omega}_2$, ie. the two-loop logarithm of the amplitude.

In principle, there are also doublet conditions where we cut certain loops $AB$ and $CD$ but it is the inequalities with another loop $EF$ -- $\la ABEF\ra$, $\la CDEF\ra$ which are violated. These cuts do exist in higher-cycle geometries, but they are harder to identify and we will not need them for our discussion here. 

\subsection{Tree numerator and double pole}

It is easy to see that for an arbitrary graph, the factorized numerator (\ref{treenum}) satisfies all singlet and doublet conditions,
\begin{equation}
\label{closed3}
    {\cal N}^{\rm tree}_G = \langle 1234\ra^{L{+}1} \times \prod_{\rm pos\,links } N_{ij}^{(+)} \times \prod_{\rm neg\, links } N_{ij}^{(-)} \,.
\end{equation}
We refer to it as the \emph{tree numerator}, as it gives the correct expression for all tree graphs. The numerator ${\cal N}^{\rm tree}_G$ is not correct for a general graph (which contains closed cycles) for one simple reason -- it fails to satisfy an important consistency condition: logarithmic singularities of the canonical form. To see this explicitly, let us consider a simplest positive geometry with one closed cycle,
\begin{center}
	\begin{tabular}{cc}
	 \includegraphics[scale=.75]{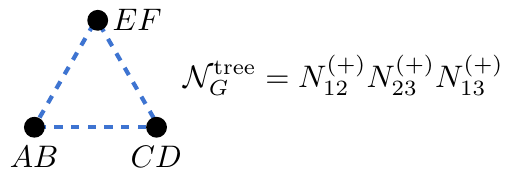}
	\end{tabular}
\end{center}
where though we chose all links to be positive, the signs are in fact irrelevant. We perform a cut when all three lines $AB$, $CD$ and $EF$ cut the line $(12)$ and in addition $AB$ intersects $CD$,
\begin{equation}
\la AB12\ra = \la CD12\ra = \la EF12\ra = \la ABCD\ra = 0\,. \label{cut1}
\end{equation}
To see the momentum twistor geometry of the solution to this cut we first let all three lines intersect line $(12)$, and then let the lines $AB$ and $CD$ intersect each other on $(12)$. This merges their intersection points and we can label them as $A=C$,
\begin{center}
	\begin{tabular}{cc}
	 \includegraphics[scale=.67]{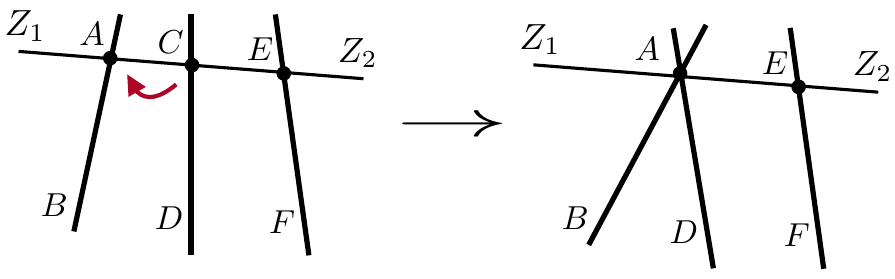}
	\end{tabular}
\end{center}
This is an allowed cut for any sign assignments, ie. the signs of $\la ABEF\ra$ and $\la CDEF\ra$ are not fixed on this cut. Now we further cut $\la ABEF\ra=0$ by merging the intersection points $A$ and $E$ on line $(12)$,
\begin{equation}
	\begin{tabular}{cc}
	 \includegraphics[scale=.67]{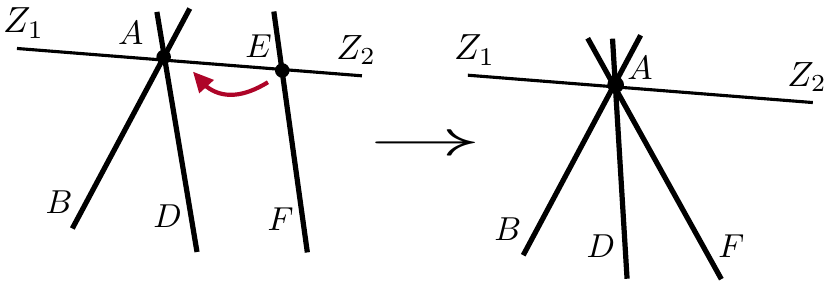}
	 \label{cutdouble}
	\end{tabular}
\end{equation}
As we can see in this configuration, there is one accidental relation
\begin{equation}
\la CDEF\ra = 0\,, \label{cut2}
\end{equation}
where lines $CD$ and $EF$ also intersect. This means that in the canonical form we generate a double pole on the cut (\ref{cutdouble})
\begin{equation}
\Omega_G \xrightarrow{cut} \frac{d\alpha\,{\cal N}_G}{\alpha^2\dots} \,,
\end{equation}
where $\alpha$ merges points $A$ and $E$ in some parametrization. Hence the numerator ${\cal N}_G$ has to vanish on the configuration (\ref{cutdouble}) in order to cancel the double pole and preserve logarithmic singularities. The tree numerator ${\cal N}^{\rm tree}_G$ fails to vanish on this cut and hence it is not the correct expression. The same applies for all other sign assignments in the triangle. 

\section{All one-cycle geometries}
\label{sec:result}

The main result of this paper is the canonical form for all positive and negative geometries with one internal loop (cycle). As we saw in the previous section, this not just given by the naive formula (\ref{treenum}). We will refer to these numerators as the ``tree part'' as this is exactly what the correct result would be if the graph was a tree. The formula (\ref{treenum}) is actually a very good start that only needs to be corrected by adding an extra ``remainder'',
\begin{equation}
{\cal N}^{1-{\rm cycle}}_G =  {\cal N}^{\rm tree}_G  + R^{\rm 1-cycle}_G\,,
\end{equation} 
where the remainder $R^{\rm 1-cycle}_G$ is yet-to-be fixed.

\subsection{Triangle geometry}

Let us go back to the positive geometry (\ref{closed3}) and fix the form from the conditions we mentioned in section \ref{sec:conditions}. We write the dlog form as 
\begin{equation}
	\begin{tabular}{cc}
	 \includegraphics[scale=.75]{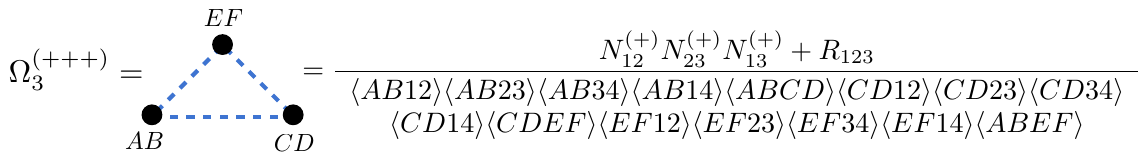}
	 \label{triang1}
	\end{tabular}
\end{equation}
We write the numerator for this sign assignment on the links as
\begin{equation}
{\cal N}^{(+++)}_{\rm triangle} = N_{12}^{(+)}N_{23}^{(+)}N_{13}^{(+)} + R_{123}\,,
\end{equation}
where the remainder $R_{123}$ is yet to be fixed. Naively, for other geometries where some of the positive links are flipped to negative, the remainder is different. For example,
\begin{equation}
	\begin{tabular}{cc}
	 \includegraphics[scale=.75]{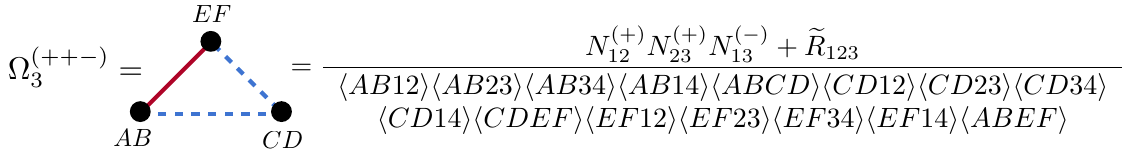}
	 \label{triang2}
	\end{tabular}
\end{equation}
The union of these two spaces has no $\la ABEF\ra$ link, hence it is a tree graph whose canonical form has a simple factorized numerator,
\begin{center}
	\begin{tabular}{cc}
	 \includegraphics[scale=.83]{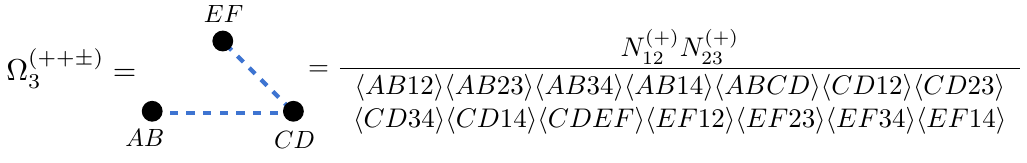}
	\end{tabular}
\end{center}
We can also get this form by summing (\ref{triang1}) and (\ref{triang2}), which puts a constraint on the one-cycle remainder terms $R_{123}$ and $\widetilde{R}_{123}$,
\begin{equation}
{\cal N}_G = N_{12}^{(+)}N_{23}^{(+)} + \frac{R_{123}+\widetilde{R}_{123}}{\la ABEF\ra\la1234\ra} = N_{12}^{(+)}N_{23}^{(+)} \,,
\end{equation}
which forces $\widetilde{R}_{123}=-R_{123}$. The same also applies to the other two assignments of signs to internal links (modulo relabeling of loops), 
\begin{center}
	\begin{tabular}{cc}
	 \includegraphics[scale=.75]{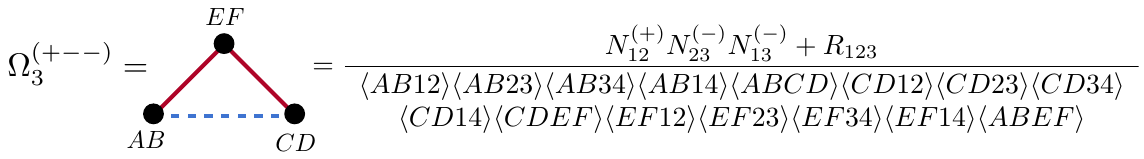}
	\end{tabular}
\end{center}
\begin{equation}
\label{eq:L3cycle1}
	\begin{tabular}{cc}
	 \includegraphics[scale=.75]{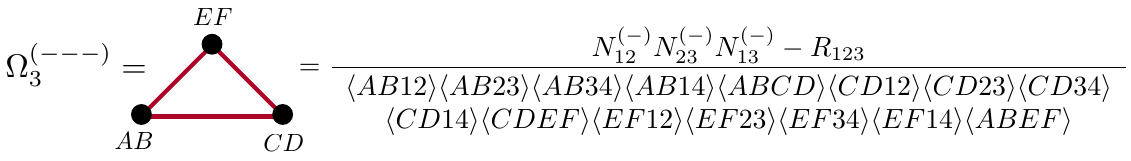}
	\end{tabular}
\end{equation}
This remainder $R_{123}$ is thus universal for all triangle geometries (up to a relative sign). We know that all these geometries suffer from the presence of double poles, and the purpose of $R_{123}$ is to cancel them. The explicit form for $R_{123}$ was already presented in \cite{Arkani-Hamed:2021iya},
\begin{equation}
R_{123} = 4M_1 - M_2^a - M_2^b - M_2^c\,, \label{rem}
\end{equation}
where we defined
\begin{align*}
M_1 =& \la AB12\ra\la AB34\ra\la CD12\ra\la CD34\ra\la EF12\ra\la EF34\ra\nonumber\\
&\hspace{5cm}+\la AB23\ra\la AB14\ra\la CD23\ra\la CD14\ra\la EF23\ra\la EF14\ra\,,\nonumber\\
M_2^a =& \la AB12\ra\la AB34\ra\times\left(\la CD12\ra \la EF34\ra + \la CD34\ra\la EF12\ra\right)\nonumber\\
&\hspace{5cm}\times\left(\la CD13\ra \la EF24\ra + \la CD24\ra\la EF13\ra\right)\nonumber\\
&+\la AB23\ra\la AB14\ra\times(\la CD23\ra \la EF14\ra + \la CD14\ra\la EF23\ra)\nonumber\\
&\hspace{5cm}\times(\la CD13\ra \la EF24\ra + \la CD24\ra\la EF13\ra)\,,
\end{align*}
where $M_2^b$ and $M_2^c$ differ from $M_2^a$ by permutation of lines $(AB),(CD)$ and $(EF)$ ($M_2^a$ is already symmetric in $(CD){\leftrightarrow}(EF)$). This can be obtained by a brute-force method: write down a complete ansatz for $R_{123}$ and impose the double pole cancellation constraint on the numerator ${\cal N}_G$. Interestingly, this fixes the form $R_{123}$ uniquely.

\subsection{Detailed construction of the remainder}

Let us now construct the remainder $R_{123}$ systematically and provide more details. We know that the tree numerators built from $N_2^{(+)}$ and $N_2^{(-)}$ automatically satisfy all singlet and doublet conditions for any graph. Therefore, the remainder $R_{123}$ must also satisfy all these conditions. But we also know that $R_{123}$ is universal for all sign assignments of links, hence it must actually vanish for \emph{all} singlet and doublet conditions: {\bf P1}, {\bf N1}, {\bf P2a}, {\bf P2b}, {\bf N2a} and {\bf N2b} for \emph{any pair} of lines $(AB),(CD)$ and $(EF)$. To build this remainder $R_{123}$ we use a particular set of building blocks 
\begin{align}
\label{eq:bb1}
N_{ij}^{(a)} &\equiv \la AB_i12\ra \la AB_j34\ra + \la AB_i34\ra\la AB_j12\ra\,,\\
\label{eq:bb2}
N_{ij}^{(b)} &\equiv \la AB_i23\ra \la AB_j14\ra + \la AB_i14\ra\la AB_j23\ra\,,\\
\label{eq:bb3}
N_{ij}^{(c)} &\equiv \la AB_i13\ra \la AB_j24\ra + \la AB_i24\ra\la AB_j13\ra\,.
\end{align}
It will also be useful to use a short-hand notation for the ``diagonal elements''
\begin{align}
\label{eq:bb4}
N_{i}^{(a)} &\equiv \frac12 N_{ii}^{(a)} = \la AB_i12\ra \la AB_i34\ra\,, \\
\label{eq:bb5}
N_{i}^{(b)} &\equiv \frac12 N_{ii}^{(b)} = \la AB_i23\ra \la AB_i14\ra\,.
\end{align}
Note that the two-loop numerators are just
\begin{equation}
N_{ij}^{(+)} = N_{ij}^{(a)} + N_{ij}^{(b)},\qquad N_{ij}^{(-)} = N_{ij}^{(c)}\,.
\end{equation}
We can then write a general ansatz in terms of these objects $N_{ij}^{(a,b,c)}$ as
\begin{equation}
R_{123} = \sum c_{k}\,N_{xy}^{(a,b,c)}N_{zw}^{(a,b,c)}N_{pq}^{(a,b,c)}\,,
\end{equation}
where $1,2$ and $3$ each appear twice in $\{x,y,z,w,p,q\}$ with some choices of $a,b,c$ superscripts, and $c_k$ are so far undetermined numerical factors. We \emph{demand} that all terms in our basis satisfy all singlet and doublet conditions \emph{individually} -- this is our assumption. Analyzing individual building blocks we learn that
\begin{enumerate}
\item $N_{ij}^{(a)}$ and $N_{ij}^{(b)}$ are designed to satisfy conditions {\bf P1} for both $i$ and $j$, and {\bf P2a} and {\bf P2b} for a link $(i,j)$.
\item $N_{ij}^{(c)}$ is designed to satisfy conditions {\bf N1} for both $i$ and $j$, and {\bf N2a} and {\bf N2b} for a link $(i,j)$.
\item Products $N_i^{(a)}$ or $N_i^{(b)}$ satisfies {\bf P1} and {\bf N1} for $i$, and also {\bf P2b}, {\bf N2a} and {\bf N2b} for any link $(i,\ast)$. 
\item Products $N_i^{(a)}N_{j\ast}^{(a)}$ and $N_i^{(b)}N_{j\ast}^{(b)}$ satisfy further condition {\bf P2a} for a link $(i,j)$. 
\item Products $N_i^{(a)}N_j^{(a)}$ and $N_i^{(b)}N_j^{(b)}$ satisfy {\bf all} conditions for both $i,j$ and the link $(i,j)$. 
\end{enumerate}
All conditions must be satisfied for all three pairs $12$, $13$ and $23$ (because our graph has all three links), and the complete basis of such monomials is (modulo relabeling of 1,2,3)
\begin{align}
\bigg\{N_1^{(a)}N_2^{(a)}N_3^{(a)},N_1^{(b)}N_2^{(b)}N_3^{(b)},N_1^{(a)}N_{23}^{(a)}N_{23}^{(c)},N_1^{(b)}N_{23}^{(b)}N_{23}^{(c)}\bigg\}\,.
\end{align}
Nothing else satisfies all constraints. Our ansatz for the remainder $R_{123}$ is then
\begin{align}
R_{123} &= c_1\cdot N_1^{(a)}N_2^{(a)}N_3^{(a)} + c_2\cdot N_1^{(b)}N_2^{(b)}N_3^{(b)}\nonumber \\
& + c_3\cdot \bigg\{N_1^{(a)} N_{23}^{(a)} N_{23}^{(c)} + N_2^{(a)} N_{13}^{(a)} N_{13}^{(c)} + N_3^{(a)}N_{12}^{(a)} N_{12}^{(c)} \bigg\} \nonumber \\
& + c_4\cdot \bigg\{N_1^{(b)}N_{23}^{(b)}N_{23}^{(c)} + N_2^{(b)}N_{13}^{(b)}N_{13}^{(c)} +N_3^{(b)}N_{12}^{(b)}N_{12}^{(c)} \bigg\}\,,\label{ansatz3}
\end{align}
and the full numerator is
\begin{align}
{\cal N}^{(+++)}_{\rm triangle} = N_{12}^{(+)}N_{23}^{(+)}N_{13}^{(+)} + R_{123}\,.
\end{align}
Now we impose the double pole cancellation condition on ${\cal N}^{(+++)}_{\rm triangle}$. This gives a unique solution which fixes the coefficients in the ansatz (\ref{ansatz3}) to be
\begin{equation}
c_1 = c_2 = 4\,, \quad c_3 = c_4 = -1\,.
\end{equation} 
Note that the cyclic symmetry (in $Z_1$, $Z_2$, $Z_3$, $Z_4$) forces us to have $c_1=c_2$ and $c_3=c_4$, but this is automatic in the solution. We can then write the result in a compact form,
\begin{equation}
R_{123} = \bigg\{ 4 N_1^{(a)}N_2^{(a)}N_3^{(a)} - \left(N_1^{(a)} N_{23}^{(a)} N_{23}^{(c)} + N_2^{(a)}N_{13}^{(a)}N_{13}^{(c)} + N_3^{(a)}N_{12}^{(a)}N_{12}^{(c)}\right)\bigg\} + (a\rightarrow b)\,,
\end{equation}
where $a\rightarrow b$ sends the ``$a$'' superscript to ``$b$'', ie. $N_i^{(a)} \rightarrow N_i^{(b)}$ and $N_{ij}^{(a)}\rightarrow N_{ij}^{(b)}$, leaving the terms with superscript ``c'' untouched. This is exactly the formula (\ref{rem}) now written in this new notation. We can easily check that the same $R_{123}$ expression trivially takes care of the double poles in the $(++-)$, $(+--)$ and $(---)$ spaces. 

As a consistency check we can verify that $\Omega^{(+++)}_{\rm triangle}$ reproduces exactly the integrand for the 3-loop amplitude $\Omega_3$, but now written in a very different way than the usual sum of double boxes and tennis court integrands. Similarly, $\Omega^{(---)}_{\rm triangle}$ together with tree graphs reproduces $\widetilde{\Omega}_3$, ie. the three-loop logarithm of the amplitude.

Note that there are actually more constraints involving the cut of all three loops than we used -- we only used the double pole cancellation condition. We do not need the other constraints, as the numerator is already fully fixed. For example, for the $(+++)$ space we can consider a cut
\begin{equation}
\la AB12\ra= \la ABCD\ra = \la CD34\ra = \la EF23\ra = \la EF14\ra = 0
\end{equation}
which is another example of a ``non-planar cut'' that is not supported by any planar diagram. 
\begin{center}
	\begin{tabular}{cc}
	 \includegraphics[scale=.64]{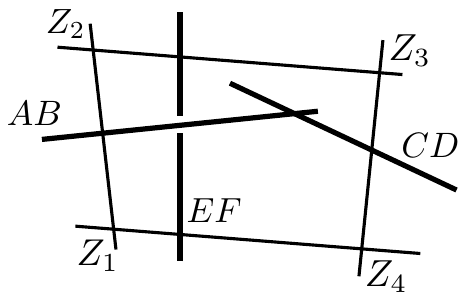}
	\end{tabular}
\end{center}
Indeed on this cut $\la ABEF\ra<0$ and $\la CDEF\ra<0$ and the numerator ${\cal N}^{(+++)}_{\rm triang}$ must vanish. This requires non-trivial cancellations in the numerator which do indeed happen. These conditions are harder to classify, so we are lucky that singlet and doublet conditions along with double pole cancellation are enough to fix the remainder $R_{123}$ completely. 

The next set of geometries we consider are polygons for any $L$. We already solved the problem for $L=3$ and before writing a general solution, we look at $L=4$ and $L=5$ examples.

\subsection{Square and pentagon geometries}

Let us now consider the next case of a simple geometry with one internal cycle, it is the $L=4$ (four loops) `square' geometry with only four links which form a closed cycle. For all links positive, we have
\begin{center}
	\begin{tabular}{cc}
	 \includegraphics[scale=.88]{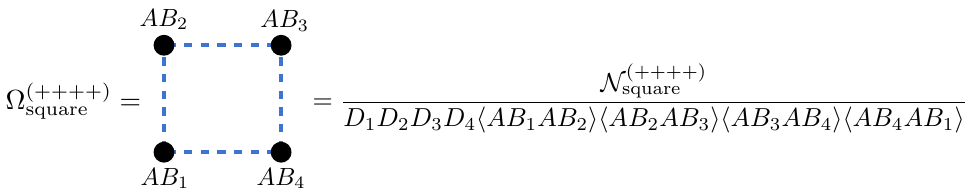}
	\end{tabular}
\end{center}
We start with writing the numerator as the sum of the tree part and the remainder, and for all links positive we have
\begin{equation}
{\cal N}^{(++++)}_{\rm square} = N_{12}^{(+)}N_{23}^{(+)}N_{34}^{(+)}N_{14}^{(+)} + R_{1234}\,. \label{square1}
\end{equation}
The remainder is again universal for any sign assignments on links. For example, if we flip the sign of the $\la AB_1AB_4\ra$ link we have the same denominator, but
\begin{equation}
{\cal N}^{(+++-)}_{\rm square} = N_{12}^{(+)}N_{23}^{(+)}N_{34}^{(+)}N_{14}^{(-)} - R_{1234}\,, \label{square2}
\end{equation}
such that the union of two geometries is a tree graph,
\begin{center}
	\begin{tabular}{cc}
	 \includegraphics[scale=.88]{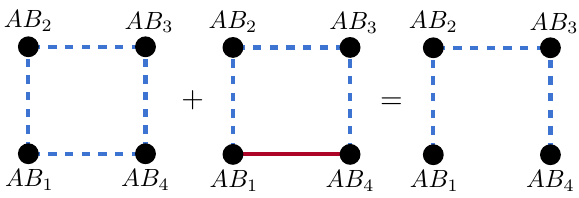}
	\end{tabular}
\end{center}
and indeed the sum of (\ref{square1}) and (\ref{square2}) reproduce the correct tree form. The remainder must satisfy all singlet and doublet conditions for the links in the graph (not all links -- since links $\la AB_1AB_3\ra$ and $\la AB_2AB_4\ra$ are missing, their signs are unrestricted) and the double pole cancellation. The double pole is again generated in the denominator of the form $\Omega_G$ if all lines $AB_i$ cut line $(12)$ (or any other external line) and intersect each other on that line -- we denote the intersection point $A$,  
\begin{center}
	\begin{tabular}{cc}
	 \includegraphics[scale=.65]{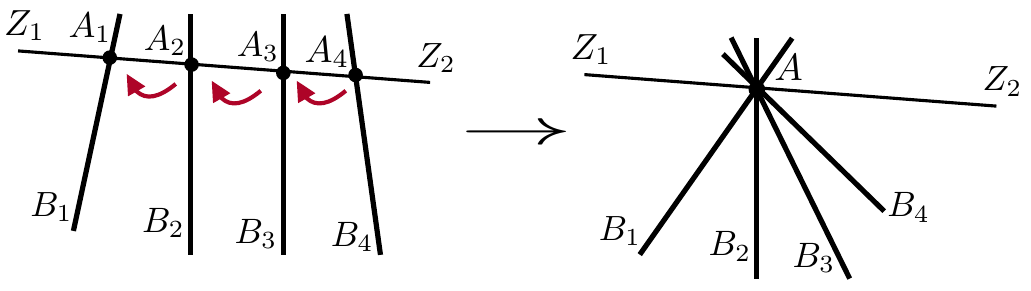}
	\end{tabular}
\end{center}
Merging the intersection points on line $(12)$ of lines $AB_1 \rightarrow AB_2 \rightarrow AB_3 \rightarrow AB_4$ also accidentally cuts the propagator $\la AB_1AB_4\ra$ which produces a double pole unless the numerator cancels it. As in the triangle case, we use the same building blocks (\ref{eq:bb1})-(\ref{eq:bb5}) and same rules for constructing the basis of terms which satisfy the singlet conditions  {\bf P1} and {\bf N1} and doublet conditions {\bf P2a}, {\bf P2b}, {\bf N2a} and {\bf N2b}. The basis now has two types of terms:
\begin{align}
N_1^{(a)}N_2^{(a)}N_3^{(a)}N_4^{(a)}\quad \mbox{and}\quad N_1^{(a)} N_2^{(a)} N_{34}^{(a)} N_{34}^{(c)}
\end{align}
Both terms cancel for all singlet conditions trivially, and also for all doublet conditions -- this is true not only for four links in the graph, but actually for all six possible links including $\la AB_1AB_3\ra$ and $\la AB_2AB_4\ra$ which are not in the graph! The cyclicity of the diagram then gives us the ansatz
\begin{align}
R_{1234} &= c_1\,N_1^{(a)}N_2^{(a)}N_3^{(a)}N_4^{(a)} + c_2 \bigg\{N_1^{(a)} N_3^{(a)} N_{24}^{(a)} N_{24}^{(c)} + N_2^{(a)} N_4^{(a)} N_{13}^{(a)} N_{13}^{(c)}\bigg\}  \label{R1234ans}\\
&\hspace{-0.5cm}+ c_3 \bigg\{N_1^{(a)} N_2^{(a)} N_{34}^{(a)} N_{34}^{(c)} {+} N_2^{(a)} N_3^{(a)} N_{14}^{(a)} N_{14}^{(c)} {+} N_3^{(a)} N_4^{(a)} N_{12}^{(a)} N_{12}^{(c)} {+} N_1^{(a)} N_4^{(a)} N_{23}^{(a)} N_{23}^{(c)} \Bigg\}\nonumber\\
& \hspace{12cm}+ (a\rightarrow b)\,. \nonumber
\end{align}
We also implicitly cycle in external legs, which translates into switching $a$ labels to $b$ labels. Note that the second and third terms in (\ref{R1234ans}) are structurally the same but the labels are distributed differently and we have to consider them separately. Imposing the condition {\bf D} on the numerator (\ref{square1}) reveals no solution. This means that our ansatz is not big enough. Indeed, there are more building blocks we can define which are analogous to $N_{ij}^{(a,b,c)}$ but now involve four lines $AB_i$ rather than just two:
\begin{align}
N^{(a)}_{1234} &= \la AB_112\ra\la AB_234\ra\la AB_312\ra\la AB_434\ra +\la AB_134\ra\la AB_212\ra\la AB_334\ra\la AB_412\ra\,,\\
N^{(b)}_{1234} &= \la AB_123\ra\la AB_214\ra\la AB_323\ra\la AB_414\ra +\la AB_114\ra\la AB_223\ra\la AB_314\ra\la AB_423\ra\,,\\
N^{(c)}_{1234} &= \la AB_113\ra\la AB_224\ra\la AB_313\ra\la AB_424\ra +\la AB_124\ra\la AB_213\ra\la AB_324\ra\la AB_413\ra\,.
\end{align}
We see an intriguing pattern of labels here which will later generalize to higher $L$. Because of the number of lines involved in the building blocks, we refer to $N_i^{(a,b)}$ as \emph{one-line invariants}, $N_{ij}^{(a,b,c)}$ as \emph{two-line invariants} and $N_{ijkl}^{(a,b,c)}$ as \emph{four-line invariants}. There are no three-line invariants and we indeed did not need any extra building blocks to solve the triangle problem.

Note that the four-line invariants are cyclically invariant (in internal lines $AB_1$, $AB_2$, $AB_3$ and $AB_4$ -- in this ordering), and do not enjoy the full permutational symmetry in $AB_j$. As for satisfying conditions {\bf P1}, {\bf N1}, {\bf P2a}, {\bf P2b}, {\bf N2a} and {\bf N2b}, they perform a similar job to $N_{ij}^{(a,b,c)}$: they do satisfy singlet conditions for all lines $AB_i$, but the doublet conditions are only satisfied for pairs $(AB_1,AB_2)$, $(AB_2,AB_3)$, $(AB_3,AB_4)$ and $(AB_4,AB_1)$. This is exactly what is needed for the topology of the graph. Hence the new building blocks to add to our ansatz are $N^{(a)}_{1234}N^{(c)}_{1234}$ and $N^{(b)}_{1234}N^{(c)}_{1234}$. Our corrected ansatz is then 
\begin{align}
R_{1234} &= c_1\,N_1^{(a)}N_2^{(a)}N_3^{(a)}N_4^{(a)} + c_2 \bigg\{N_1^{(a)} N_3^{(a)} N_{24}^{(a)} N_{24}^{(c)} + N_2^{(a)} N_4^{(a)} N_{13}^{(a)} N_{13}^{(c)}\bigg\}  \label{R1234ans2}\\
&\hspace{-0.5cm}+ c_3 \bigg\{N_1^{(a)} N_2^{(a)} N_{34}^{(a)} N_{34}^{(c)} {+} N_2^{(a)} N_3^{(a)} N_{14}^{(a)} N_{14}^{(c)} {+} N_3^{(a)} N_4^{(a)} N_{12}^{(a)} N_{12}^{(c)} {+} N_1^{(a)} N_4^{(a)} N_{23}^{(a)} N_{23}^{(c)} \Bigg\}\nonumber\\
&\hspace{-0.5cm}+ c_4\, N^{(a)}_{1234}N^{(c)}_{1234}\hspace{9.8cm} + (a\rightarrow b)\,. \nonumber
\end{align}
Imposing the double pole cancellation condition on the full numerator ${\cal N}^{(++++)}_{\rm square}$, we find an unique solution,
\begin{equation}
c_1 = 12, \quad c_2=c_3=c_4=-1\,.
\end{equation}
We can now also check many other cuts which involve more than two lines where the numerator should vanish and find always a perfect match. Note that going from one sign assignment to another changes the sign of the remainder, each flip from positive to negative changes the sign of $R_{1234}$.

Let us do one more explicit example before writing down the solution for a general polygon. Let us consider $L=5$ which is a pentagon,
\begin{center}
	\begin{tabular}{cc}
	 \includegraphics[scale=.78]{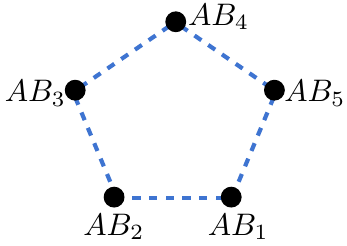}
	\end{tabular}
\end{center}
The ansatz for the numerator of a geometry with all links positive is 
\begin{equation}
{\cal N}^{(+++++)}_{\rm pentagon} = N^{(+)}_{12}N^{(+)}_{23}N^{(+)}_{34}N^{(+)}_{45}N^{(+)}_{15} + R_{12345}\,,
\end{equation}
where the ansatz for the remainder is now 
\begin{align}
R_{12345} &= c_1\,N_1^{(a)}N_2^{(a)}N_3^{(a)}N_4^{(a)}N_5^{(a)} + c_2 \bigg\{N_1^{(a)} N_2^{(a)} N_3^{(a)} N_{45}^{(a)} N_{45}^{(c)} + {\rm cycl}\bigg\}  \label{R12345ans}\\
&\hspace{-0.5cm} + c_3 \bigg\{N_1^{(a)} N_2^{(a)} N_4^{(a)}N_{35}^{(a)} N_{35}^{(c)} + {\rm cycl}\Bigg\}
+ c_4 \bigg\{N_1^{(a)} N_{2345}^{(a)}N_{2345}^{(c)} + {\rm cycl}\Bigg\} \quad + (a\rightarrow b)\,, \nonumber
\end{align}
where the `cycl' refers to cycling in the loop lines $AB_i\rightarrow AB_{i{+}1}$. Note that the ansatz does not involve any five-line invariants. The dangerous configuration that produces the double pole in the canonical form is similar to the previous cases,
\begin{center}
	\begin{tabular}{cc}
	 \includegraphics[scale=.7]{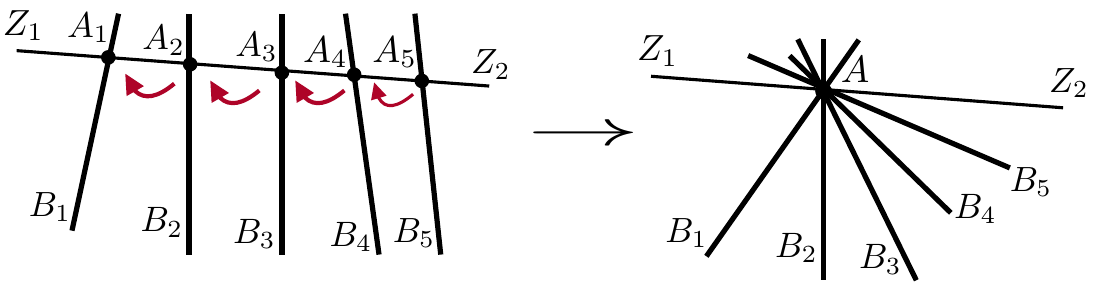}
	\end{tabular}
\end{center}
and we again find a unique solution, 
\begin{equation}
c_1 = 28, \quad c_2=c_3=c_4=-1\,.
\end{equation}
We can now check many additional cuts which are illegal in a given positive geometry and verify that our numerator vanishes where it should. For example, one `planarity condition' is the following cut:
\begin{equation*}
\la AB_1 12\ra = \la AB_1 AB_2 \ra = \la AB_2 AB_3\ra = \la AB_3 34 \ra = \la AB_4 23\ra = \la AB_4 AB_5 \ra = \la AB_514\ra = 0
\end{equation*}
\begin{center}
	\begin{tabular}{cc}
	 \includegraphics[scale=.7]{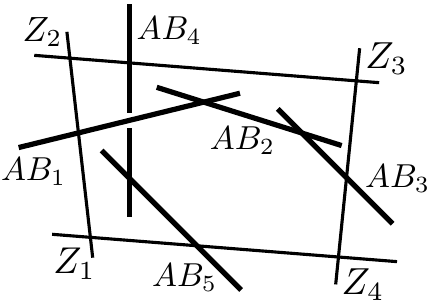}
	\end{tabular}
\end{center}
Very non-trivially, the numerator ${\cal N}^{(+++++)}_{\rm pentagon}$ vanishes on this cut. Flipping a sign of a link again changes $N^{(+)}_{ij} \rightarrow N^{(-)}_{ij}$ in the tree part of the numerator and flips the sign of the remainder $R_{12345}$. For the totally negative geometry (with all links negative) we have
\begin{equation}
{\cal N}^{(-----)}_{\rm pentagon} = N^{(-)}_{12}N^{(-)}_{23}N^{(-)}_{34}N^{(-)}_{45}N^{(-)}_{15} - R_{12345}\,.
\end{equation}
%

\subsection{General one-cycle graph}

Let us now consider a general polygon geometry at $L$-loops (with $L$ vertices). For all links positive 
\begin{center}
	\begin{tabular}{cc}
	 \includegraphics[scale=.75]{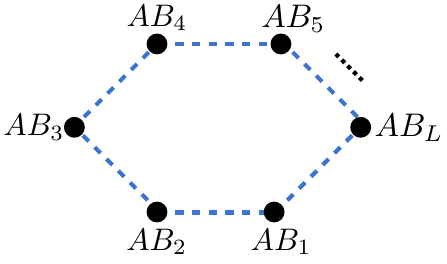}
	\end{tabular}
\end{center}
the numerator of the form is
\begin{equation}
{\cal N}_G = \prod_{i=1}^L N^{(+)}_{i,i{+}1} + R^{\rm 1-loop}_L\,,
\end{equation}
where the remainder $R^{\rm 1-loop}$ takes this form
\begin{equation}
\label{eq:R1Loop}
\boxed{R^{\rm 1-loop}_L = \bigg\{(2^L{-}4)N_{1}^{(a)}N_{2}^{(a)} \cdots N_{L}^{(a)} {-} \sum_{p\in {\rm even}}\sum_{i} \big\{N_{i_1i_2{\dots}i_p}^{(a)}N_{i_1i_2{\dots}i_p}^{(c)}N_{i_p{+}1}^{(a)}\cdots N_{i_L}^{(a)}\big\}\bigg\} {+} (a{\rightarrow} b)}
\end{equation}
We use the superscript `1-loop' here and later denoting that the object is associated with the cycle (loop) in the diagram. The sum in (\ref{eq:R1Loop}) is over orderings $i$ are all selections of $p$ labels where the $p$ labels are ordered within the argument of the factors $N_{i_1i_2{\dots}i_p}^{(a,b,c)}$. These are defined as
\begin{align}
N_{12\dots p}^{(a)} &= \prod_{j\in {\rm odd}} \la AB_j12\ra \prod_{k\in {\rm even}}\la AB_k34\ra + \prod_{j\in {\rm odd}} \la AB_j34\ra \prod_{k\in {\rm even}}\la AB_k12\ra\\
N_{12\dots p}^{(b)} &= \prod_{j\in {\rm odd}} \la AB_j23\ra \prod_{k\in {\rm even}}\la AB_k14\ra + \prod_{j\in {\rm odd}} \la AB_j14\ra \prod_{k\in {\rm even}}\la AB_k23\ra\\
N_{12\dots p}^{(c)} &= \prod_{j\in {\rm odd}} \la AB_j13\ra \prod_{k\in {\rm even}}\la AB_k24\ra + \prod_{j\in {\rm odd}} \la AB_j24\ra \prod_{k\in {\rm even}}\la AB_k13\ra
\end{align}
For example, for the $L=6$ hexagon we have
\begin{align}
R^{\rm 1-loop}_{123456}&= \bigg\{60N_1^{(a)}N_2^{(a)}N_3^{(a)}N_4^{(a)}N_5^{(a)}N_6^{(a)} - \big\{N_1^{(a)}N_2^{(a)}N_3^{(a)}N_4^{(a)}N_{56}^{(a)}N_{56}^{(c)}+{\rm perm}\big\}  \nonumber \\
& \hspace{1cm} - \big\{N_1^{(a)}N_2^{(a)}N_{3456}^{(a)}N_{3456}^{(c)} + {\rm perm}\big\} - N_{123456}^{(a)}N_{123456}^{(c)}\bigg\} + (a{\rightarrow}b) \label{hexagon}
\end{align}
Note in the second and third terms the sums are over all permutations of lines $AB_i$ (we only take distinct terms, no duplicates), only the last fourth term does not respect permutational symmetry and it is only cyclically invariant. For the totally negative polygon we have 
\begin{center}
	\begin{tabular}{cc}
	 \includegraphics[scale=.7]{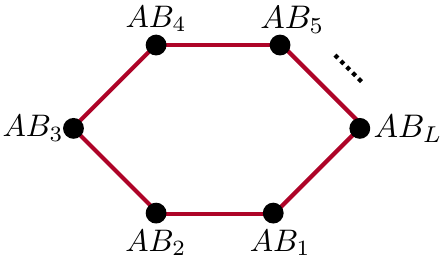}
	\end{tabular}
\end{center}
\begin{equation}
\label{eq:result1Loop}
{\cal N}_G = \prod_{i=1}^L N^{(-)}_{i,i{+}1} + (-1)^{L}R^{\rm 1-loop}_L
\end{equation}
Finally, let us consider a completely generic one-loop graph, ie. connected graph with one cycle (here for negative links only). There is always a one-loop core and the tree-level branches attached to it. The denominator is given by all the one-loop factors $D_i$ and all internal propagators $\la AB_iAB_j\ra$ corresponding to the links as for any graph. The numerator can be organized as follows:
\begin{equation}
\label{eq:final1Loop}
{\cal N}_G = \prod_{\rm links} N_{ij}^{(-)} + (-1)^{L'}\prod_{\rm branches} N_{ij}^{(-)} \times R^{\rm 1-loop}_{L'}
\end{equation}
where the remainder $R^{\rm 1-loop}_{L'}$ is now for the $L'$-loop core with $AB_i$ in that particular ordering. We can simply start from the numerator for the core and multiply it by $N^{(-)}_{ij}$ for any extra link attached to it. For an explicit $L=12$ example we have (using just label $k$ for $AB_k$ for simplicity),
\begin{equation}
	\begin{tabular}{cc}
	 \includegraphics[scale=.55]{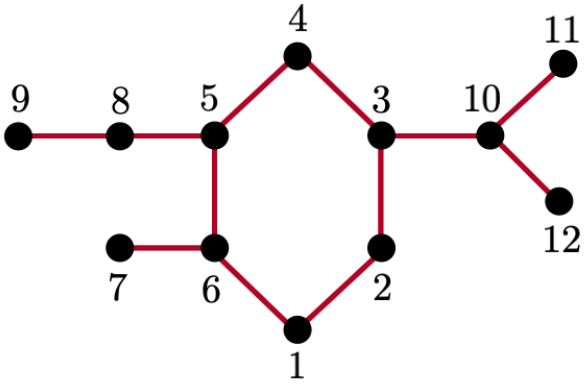} \label{four11}
	\end{tabular} 
\end{equation}
The numerator is now
\begin{align}
{\cal N}_G &= N_{67}^{(-)}N_{58}^{(-)}N_{89}^{(-)}N_{3,10}^{(-)}N_{10,11}^{(-)}N_{10,12}^{(-)} \times \bigg\{N_{12}^{(-)}N_{23}^{(-)}N_{34}^{(-)}N_{45}^{(-)}N_{56}^{(-)}N_{61}^{(-)} + R^{\rm 1-loop}_{123456}\bigg\}
\end{align}
where the loop function $R^{\rm 1-loop}_{123456}$ was given in (\ref{hexagon}). The rules for the graph with both positive and negative links is again the same up to a sign of $R^{\rm 1-loop}$. This completes a construction of a canonical form for an arbitrary graph with one internal cycle with any assignment of positive and negative links.

\section{Canonical form decomposition}
\label{sec:form}

As evident from (\ref{four2}) even for $L=4$ to get the complete result for $\widetilde{\Omega}$ we need to calculate the canonical forms for higher-loop geometries (graphs with more than one cycles). While the general principle still should work: write down the ansatz and impose vanishing on illegal cuts as constraints, it is much harder to find a similar organizational principle to the one that allowed us to solve the one-loop geometries. Our approach of loops of loops expansion is very effective in solving infinite class of graphs (and provide an approximation of the full result), but if we want an exact result at a fixed loop order $L$, we can not avoid writing down the form for a complete graph (with all $\la AB_i AB_j\ra$ links) which is then as hard as calculating the integrand of the original amplitude $\Omega_L$ (which is just one graph with all positive links)

Our approach to attack the problem of higher-loop graphs (with more than one internal cycle) is a certain \emph{decomposition} which reduces the problem of finding the canonical form to a purely higher-loop term (in the same sense as the $R^{\rm 1-loop}_G$ was the one-loop piece while the product of $N_{ij}^{(\pm)}$ was a tree piece. This approach separates `easy' or `known' terms from a truly new kind of contribution we have to calculate separately. Let us see the logic of this decomposition in the example from (\ref{four2}), and start solving the canonical form for the geometry with two cycles -- this is the only contributing term to $\widetilde{\Omega}_4^{\rm 2-loop}$. 
\begin{equation}
\label{eq:L4cycle1}
	\begin{tabular}{cc}
	 \includegraphics[scale=.77]{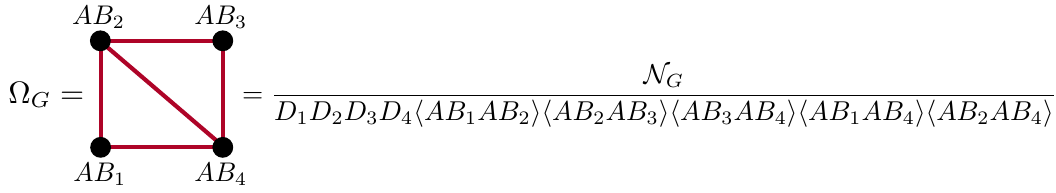}
	\end{tabular} 
\end{equation}
The numerator can be decomposed in this way,
\begin{align}
{\cal N}_G = &N_{12}^{(-)}N_{23}^{(-)}N_{34}^{(-)}N_{14}^{(-)}N_{24}^{(-)}  \nonumber\\
&\hspace{1.5cm}- R_{124}^{\rm 1-loop} N_{23}^{(-)} N_{34}^{(-)} - R_{234}^{\rm 1-loop} N_{12}^{(-)} N_{23}^{(-)}+ R_{1234}^{\rm 1-loop} N_{24}^{(-)} - R_{1234,13}^{\rm 2-loop} \label{twoloopdec}
\end{align}
We can introduce a graphic decomposion 
\begin{equation}
\label{eq:cycledecomp4}
	\begin{tabular}{cc}
	 \includegraphics[scale=.73]{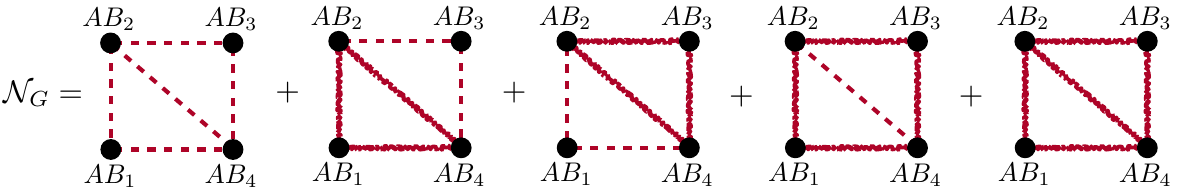}
	\end{tabular} 
\end{equation}
where the thick lines compose irreducible closed cycles (one-cycle, two-cycle) and the dashed lines are additional tree links. The minus signs in (\ref{twoloopdec}) follow the same logic as in the one-cycle geometry case (terms with an even number of negative links has relative plus sign and odd number of negative links leads to a minus sign). The last term in (\ref{eq:cycledecomp4}) is an irreducible two-loop (two-cycle) term $R^{\rm 2-loop}_{1234,13}$ which is universal for all sign assignments. This purely two-loop factor must satisfy an important constraint 
\begin{equation}
\boxed{R^{\rm 2-loop}_{1234,13} = 0 \quad \mbox{on any cut when $\la AB_i AB_j\ra$ has a definite sign}}
\end{equation}
for all the links in the two-loop graph. This is the same universality argument we used in the discussion of the one-loop function $R^{\rm 1-loop}$.

This decomposition of ${\cal N}_G$ in terms of tree, one-loop and two-loop parts makes various properties of the canonical form manifest. If we consider a totally negative geometry (\ref{eq:L4cycle1}) then if we localize one of the lines $AB_i$ on the leading singularity $AB_i=13$ or $AB_i=24$, the geometry reduces to something simpler which is also built in the form. For example, for $AB_3=13$ the geometry reduces to
\begin{equation}
	\begin{tabular}{cc}
	 \includegraphics[scale=.8]{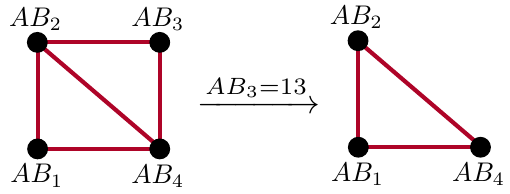}
	\end{tabular} 
\end{equation}
In the canonical form all terms with label $3$ going away: the two-loop numerators together with the mutual pole reduce to constant
\begin{equation}
\frac{N_{3k}^{(-)}}{\la AB_3 AB_k\ra} = \frac{\la AB_3 13\ra\la AB_k24\ra + \la AB_3 24\ra\la AB_k 13\ra}{\la AB_3 AB_k\ra} \xrightarrow{AB_3=13} \la 1234\ra  \label{FormRed}
\end{equation}
while any one-loop and two-loop functions involving label $3$ vanishes,
\begin{equation}
R^{\rm 1-loop}_{134} = R^{\rm 1-loop}_{1234} = R^{\rm 2-loop}_{1234,13} = 0
\end{equation}
As a result the canonical form reduces to
\begin{equation}
\Omega_G \xrightarrow{AB_3=13} \Omega_{G_{\rm red}} = \frac{N_{12}^{(-)}N_{24}^{(-)}N_{14}^{(-)} - R_{124}^{\rm 1-loop}}{D_1D_2D_4 \la AB_1AB_2\ra\la AB_1AB_4\ra\la AB_2AB_4\ra}
\end{equation}
which is exactly the correct form of the reduced graph (\ref{eq:L3cycle1}). Note that if instead of fixing $AB_3=13$ we fix now another loop line $AB_4=13$, the result is different but again correct. The graph reduces to
\begin{equation}
	\begin{tabular}{cc}
	 \includegraphics[scale=.8]{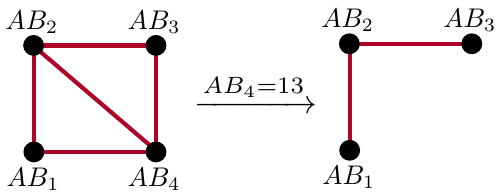}
	\end{tabular} 
\end{equation}
We get the same relation (\ref{FormRed}) now for $AB_4$, but now all one-loop and two-loop functions are zero
\begin{equation}
R^{\rm 1-loop}_{134} = R^{\rm 1-loop}_{124} = R^{\rm 1-loop}_{1234} = R^{\rm 2-loop}_{1234,13} = 0
\end{equation}
and the canonical form reduces just to the tree form,
\begin{equation}
\label{eq:reducetotree}
\Omega_G \xrightarrow{AB_4=13} \Omega_{G_{\rm red}'} = \frac{N_{12}^{(-)}N_{23}^{(-)}N_{13}^{(-)}}{D_1D_2D_3 \la AB_1AB_2\ra\la AB_1AB_3\ra\la AB_2AB_3\ra}
\end{equation}
This is precisely correct because (\ref{eq:reducetotree}) is a tree graph. The analogous procedure can be also applied to collinear limit cuts, which survive for positive links and such decomposition correctly reduces to the canonical form of a simpler graph. Another check we can perform is the sum rule, where we take a union of two geometries which differ by a sign of one link, and sum their canonical forms. For example, the analogous graph where we flip the sign of $\la AB_3AB_4\ra$ link to positive, is
\begin{equation}
\label{eq:L4cycle1pos}
	\begin{tabular}{cc}
	 \includegraphics[scale=.8]{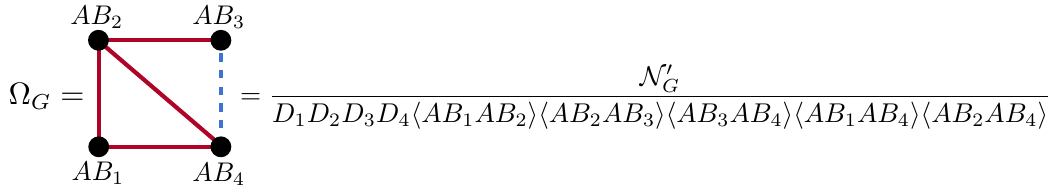}
	\end{tabular} 
\end{equation}
has an analogous decomposition
\begin{align}
{\cal N}_G' = &N_{12}^{(-)}N_{23}^{(-)}N_{34}^{(+)}N_{14}^{(-)}N_{24}^{(-)} \nonumber\\
&\hspace{1.5cm}- R_{124}^{\rm 1-loop} N_{23}^{(-)} N_{34}^{(+)} + R_{134}^{\rm 1-loop} N_{12}^{(-)} N_{23}^{(-)} - R_{1234}^{\rm 1-loop} N_{24}^{(-)} + R_{1234,13}^{\rm 2-loop} \label{twoloopdec2}
\end{align}
We can again see a geometric sum rule in action here: as the two geometries (\ref{eq:L4cycle1}) and (\ref{eq:L4cycle1pos}) only differ by a single sign of one link, we can take a union which gives us a one-loop geometry,
\begin{equation}
	\begin{tabular}{cc}
	 \includegraphics[scale=.8]{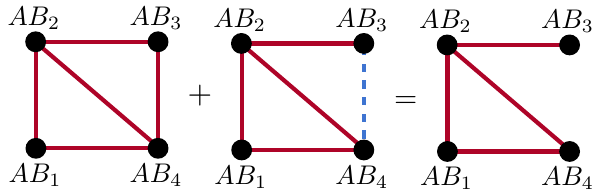}
	\end{tabular} 
\end{equation}
At the level of numerators, we can sum two terms (\ref{twoloopdec}) and (\ref{twoloopdec2}) and generate an inverse propagator $\la AB_3 AB_4\ra$ and the numerator ${\cal N}_G''$ of the new one-loop graph,
\begin{equation}
{\cal N}_G + {\cal N}_G' = \underbrace{\left(N_{34}^{(+)} + N_{34}^{(-)}\right)}_{\la AB_3AB_4\ra\la1234\ra} \times \underbrace{\left(N_{12}^{(-)}N_{23}^{(-)}N_{14}^{(-)}N_{24}^{(-)} - R_{124}^{\rm 1-loop} N_{23}^{(-)} \right)}_{{\cal N}_G''}\end{equation}
We see that the whole problem reduces to the calculation of $R_{1234,13}^{\rm 2-loop}$. The same decomposition extends to any two-loop diagram. For example, one contribution to $\widetilde{\Omega}^{\rm 2-loop}_6$ is a following negative geometry,
\begin{equation}
	\begin{tabular}{cc}
	 \includegraphics[scale=.63]{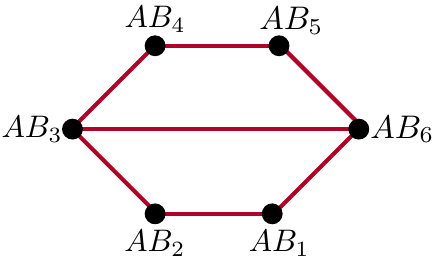}
	\end{tabular} 
\end{equation}
for which the numerator can be decomposed as
\begin{align}
{\cal N}_G &= N_{12}^{(-)}N_{23}^{(-)}N_{34}^{(+)}N_{45}^{(-)}N_{56}^{(-)}N_{16}^{(-)}N_{36}^{(-)} {+} R_{1234}^{\rm 1-loop}N_{45}^{(-)}N_{56}^{(-)}N_{16}^{(-)} {+} R_{3456}^{\rm 1-loop} N_{16}^{(-)} N_{12}^{(-)}N_{23}^{(-)} \nonumber\\ &\hspace{0.3cm} + R_{123456}^{\rm 1-loop}N_{36}^{(-)} + R_{123456,36}^{\rm 2-loop}
\end{align}
where all relative signs are $+$ because the cycles have even number of vertices each. The same decomposition also applies for canonical forms of graphs with more than two cycles. 

Now assuming we know how to find the two-loop remainder $R^{\rm 2-loop}$, we can perform this decomposition for the three-loop graph, for example the only one in $\widetilde{\Omega}_4^{\rm 3-loop}$ and the last piece needed to fully calculate $\widetilde{\Omega}_4$.

\vspace{-0.2cm}

\begin{equation}
	\begin{tabular}{cc}
	 \includegraphics[scale=.8]{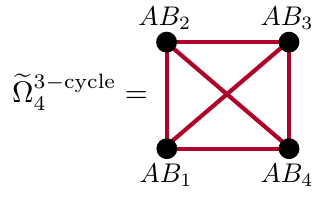}
	\end{tabular} 
\end{equation}

\vspace{-0.2cm}

As we already know, once we determine this canonical form, we can also easily calculate the canonical form $\Omega_4$ for the $L=4$ amplitude which graph has only positive links. The decomposition of the canonical form into tree, one-loop, two-loop and three-loop terms is:
\begin{align}
\Omega_G &= N_{12}^{(-)}N_{23}^{(-)}N_{34}^{(-)}N_{14}^{(-)}N_{24}^{(-)}N_{13}^{(-)}  - \sum_{\pi_1} R_{123}^{\rm 1-loop} N_{14}^{(-)} N_{24}^{(-)}N_{34}^{(-)} + \sum_{\pi_2} R_{1234}^{\rm 1-loop} N_{13}^{(-)}N_{24}^{(-)}\nonumber\\ 
&\hspace{0.3cm} - \sum_{\pi_3} R_{1234,13}^{\rm 2-loop}N_{24}^{(-)} + R_{1234}^{\rm 3-loop}
\end{align}
\begin{equation}
	\begin{tabular}{cc}
	 \includegraphics[scale=.75]{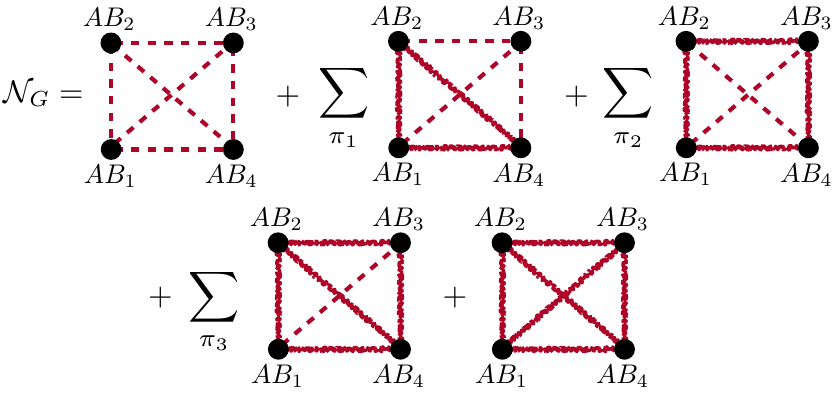}
	\end{tabular} 
\end{equation}
The permutation $\pi_1$ is over all embedded one-loop triangles, there are 4 of them $(123)$, $(124)$, $(134)$, $(234)$. The permutation $\pi_2$ is over all embedded one-loop squares (with a particular ordering), there are 3 of them $(1234)$, $(1243)$, $(1324)$. The permutation $\pi_3$ is over all embedded two-loop subtopologies (where we ignore one of the links), this is 6 terms (ignoring one of the links each). The final purely three-loop piece is again universal for any link sign assignment and satisfies
\begin{equation}
\boxed{R^{\rm 3-loop}_{1234} = 0 \quad \mbox{on any cut when $\la AB_i AB_j\ra$ has a definite sign}}
\end{equation}

\vspace{-0.3cm}
Now this is true for all links as the graph as well as $R^{\rm 3-loop}_{1234}$ is permutationally invariant in all four loop lines $AB_i$. The same logic can be used for higher loops at any $L$. 

Our procedure seems reminiscent of `Feynman rules' and 'Feynman diagrams' but now in the loops of loops space -- where the role of external particles are played by loop lines $AB_i$, the functions $N^{(\pm)}_{ij}$ are free propogators and $R^{\ell{ \rm -loop}}$ are some kind of interactions vertices. The connection is very vague, and obviously the rules how to construct our graphs are quite different, but some analogy can be found here. In either case, the problem of finding the canonical form $\Omega_G$ for a given graph is reduced to finding these $\ell$-loop irreducible functions (remainders) $R^{\ell{\rm -loop}}$.

\section{From canonical forms to polylogarithms}
\label{sec:polylogs}

So far we have studied the canonical forms for positive and negative geometries, and showed that certain sets of them reproduce the integrands for the $L$-loop amplitude $\Omega_L$ and the logarithm of the $L$-loop amplitude $\widetilde{\Omega}_L$. When integrating $\Omega_L$ or $\widetilde{\Omega}_L$ or canonical forms for individual geometries, we encounter IR divergencies which need to be regulated.

As was pointed in \cite{Arkani-Hamed:2021iya}, for negative geometries i.e. graphs with only negative links, we can define an \emph{IR finite} object by freezing one of the loops and integrating over all others. The frozen loop is marked by a cross, as shown below:
\begin{equation}
	\begin{tabular}{cc}
	 \includegraphics[scale=.75]{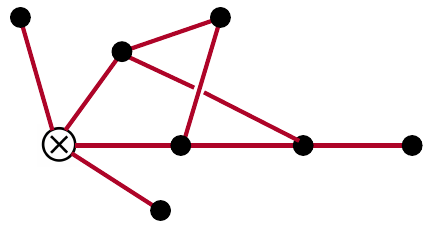}
	\end{tabular} 
\end{equation}
This is all based on the integrand-level statement about the presence of IR divergencies: in order to generate an IR divergence during integration, the integrand needs to have a singularity when one or more loop momenta access the collinear region. In the momentum twistor space, this collinear region corresponds to the line $AB$ being in a collinear configuration in the plane $(i{-}1\,i\,i{+}1)$ and passing through vertex $Z_i$. We have discussed this configuration multiple times in this paper, because in this position for $AB$, any mutual bracket $\la ABCD\ra>0$. Hence it is an illegal configuration if the vertex $AB$ in the graph is connected to any other vertex $CD$ -- which it definitely is in any connected graph. 

The only way around this is to place all loop lines $AB_i$, where none of the loops are left to violate the mutual negativity conditions. Hence freezing one of the loops ensures that the result after integration is IR finite. This was discussed in greater detail in \cite{Arkani-Hamed:2021iya}. 

\subsection{Differential equation method}

The special form of the tree numerator allows us to find a powerful differential equation. Suppose the frozen loop $AB$ is connected to the rest of the graph through a link to another loop $CD$. Then we can write for the whole integral 
\begin{equation}
    I(z) = \int_{CD} \frac{\ang{AB13}\ang{CD24}\ang{1234}}{\ang{CD12}\ang{CD23}\ang{CD34}\ang{CD41}\ang{ABCD}} H(z_{CD})\,,
\end{equation}
where $H(z_{CD})$ is the rest of the diagram integrated with the loop $CD$ frozen, with $z_{CD} = \la CD12\ra\la CD34\ra/ \la CD14\ra\la CD23\ra$. We can now act with the Laplacian $(z\partial_z)^2$ on the integral $I(z)$ and get 
\begin{equation}
    (z \partial_z)^2 I(z) = -H(z)\,.
\end{equation}
Effectively, the action of the operator reduces the graph to a simpler one with the link $(ABCD)$ removed. In the dual variables, this is 
\begin{equation}
	\begin{tabular}{cc}
	 \includegraphics[scale=.77]{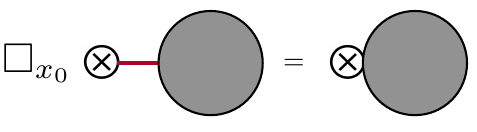}
	\end{tabular} 
\end{equation}
Not only does this work for any tree graph where the frozen loop $AB$ is connected to the rest of the graph through one link, but it can also be used for the resummation of all ladder graphs,
\begin{equation}
	\begin{tabular}{cc}
	 \includegraphics[scale=.56]{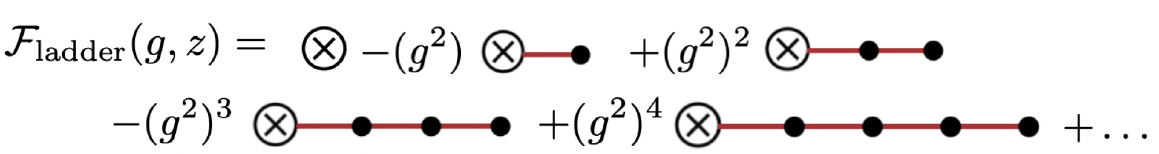}
	\end{tabular} 
\end{equation}
Imposing the boundary condition that the result must vanish for $z=-1$ -- which is just a simple consequence of the vanishing of the integrand for $\la AB13\ra = \la AB24\ra = 0$, we get \cite{Arkani-Hamed:2021iya}
\begin{equation}
    {\cal F}_{\rm ladder}=\frac{\cos{(\sqrt{2}g\log{z})}}{\cosh{(\sqrt{2}g \pi)}} \,.
\end{equation}
However, this is not a consistent approximation in our setup since we need to sum over all tree graphs to get the `leading order' contribution in the loops of loops expansion. 
\begin{equation}
	\begin{tabular}{cc}
	 \includegraphics[scale=.67]{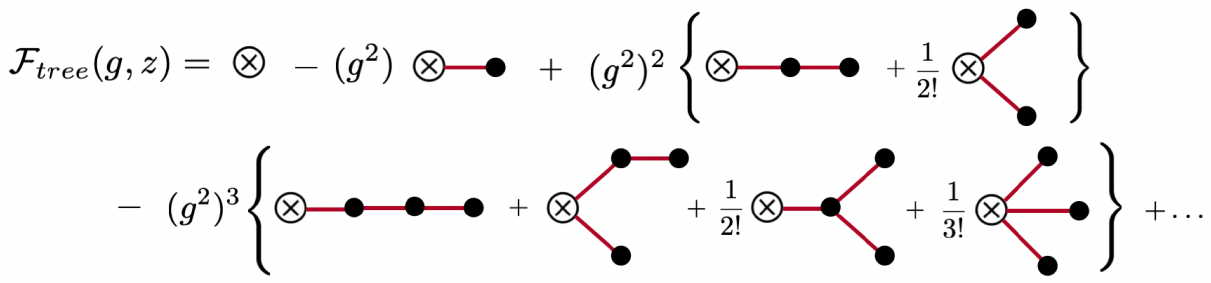}
	\end{tabular} 
\end{equation}
For a general tree graph the marked point can be somewhere inside and connected to other vertices through multiple links. To resolve that we realize that the sum of all tree graphs satisfies a simple exponentiation,
\begin{equation}
    {\cal F}_{\rm tree}(g,z) = e^{{\cal H}_{\rm tree}(g)}\,,
\end{equation}
where ${\cal F}_{\rm tree}(g,z)$ and ${\cal H}_{\rm tree}(g)$ differ in the way that marked point $AB$ can appear,
\begin{equation}
	\begin{tabular}{cc}
	 \includegraphics[scale=.55]{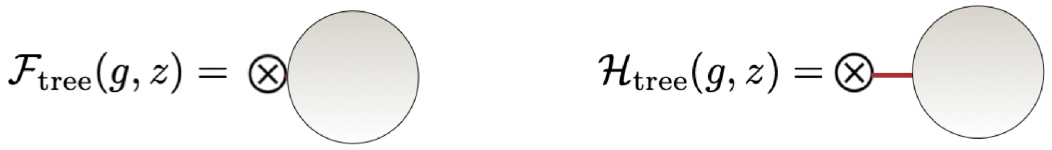}
	\end{tabular} 
\end{equation}

\vspace{-0.3cm}
and the generating functional ${\cal H}_{\rm tree}(g,z)$ now satisfies the same Laplace equation because the marked point $AB$ is now connected through a single link to the rest of the graph,
\begin{equation}
    \frac{1}{2}(z\partial_z)^2 {\cal H}_{tree}+g^2 e^{{\cal H}_{tree}}=0\,.
\end{equation}
This was solved in \cite{Arkani-Hamed:2021iya} by
\begin{equation}
    {\cal F}_{tree}(g,z)= \frac{A^2 z^A}{g^2(z^A+1)^2}
\end{equation}
with $\frac{A}{2g \cos{\pi A/2}}=1$. The differential equation results can be used to obtain ${\cal F}^{n}_{\rm tree}(g,z)$ for any order $n$ by asymptotically expanding the solution in powers of g. The refer the reader to \cite{Arkani-Hamed:2021iya} for more details. It is important to note that the Laplacian will collapse the propagator connecting the frozen loop to rest of the diagram even if the rest is not a tree. However, this is not true if the unintegrated loop is attached to multiple propagators, such as the frozen corner of the 3-loop triangle, since the boxing trick described earlier no longer works. One needs to search for new types of differential operators, perhaps analogous to \cite{Drummond:2010cz} where the operators acted on the external momentum twistors, or using methods in \cite{Henn:2023pkc}.

\subsection{New triangle formulas}

While we can not find the integrated expressions for all one cycle graphs, we still make a modest progress in this direction starting from the simplest triangle topology

\vspace{-0.2cm}

\begin{equation}
	\begin{tabular}{cc}
	 \includegraphics[scale=.75]{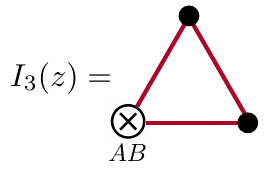}
	\end{tabular} 
\end{equation}

\vspace{-0.3cm}

Integrating over two of the loops in this diagram and leaving one of them un-integrated, or ``frozen'' as shown above, was expressed as a function of the cross-ratio $z$ in \cite{Arkani-Hamed:2021iya} as:
\begin{align}
I_3(z)  = &\frac{\pi^4}{9}+\frac{1}{3}\log^4{(z)}-2 \log^2{(z)} \left[-\frac{2}{3} {\rm Li}_2 \left(\frac{1}{1{+}z} \right)-\frac{2}{3} {\rm Li}_2 \left(\frac{z}{1{+}z} \right) + \frac{\pi^2}{9} \right] \\
 & -2 \log{(z)} \left[4 {\rm Li}_3 \left(\frac{1}{1{+}z} \right)-4 {\rm Li}_3\left(\frac{z}{1{+}z} \right) \right]+\frac{4}{3}\left[{\rm Li}_2 \left(\frac{1}{1{+}z} \right)+ {\rm Li}_2 \left(\frac{z}{1{+}z} \right) -\frac{\pi^2}{6}\right]^2 \nonumber \\
 & + \frac{16 \pi^2}{3}\left[{\rm Li}_2 \left(\frac{1}{1{+}z} \right)+ {\rm Li}_2 \left(\frac{z}{1{+}z} \right) -\frac{\pi^2}{6}\right]+16 \left[{\rm Li}_4 \left(\frac{1}{1{+}z} \right)+ {\rm Li}_4 \left(\frac{z}{1{+}z} \right) \right]\,\nonumber
\end{align}
or written in terms of harmonic polylogarithms as
\begin{align}
      I_3(z) &= 8H_{0,0,0,0}(z)+8H_{{-}1,0,0,0}(z){-}16H_{{-}1,{-}1,0,0}(z)+8H_{{-}2,0,0}(z)-8 \zeta_3 (2H_{{-}1}(z)-H_0(z)) \nonumber\\
      & +4 \pi^2(H_{{-}1,0}(z)-H_{{-}1,{-}1}(z)+H_{{-}2}(z))+\frac{13 \pi^4}{45}\,. \label{I3z}
\end{align}

Unlike all formulas for tree graphs, this expression does not satisfy the boundary condition for $z=-1$, as the integrand does not vanish in this case. Also, we do not know if there is any differential operator which would anihilate this formula. Ideally, we would like to evaluate integrals for all graphs with one cycles, but the lack of differential equation and boundary conditions does not allow us to do it at the moment.

Instead, we consider an easier problem: we fix the triangle as the cycle topology in the graph and attach one tree branch to it that contains the marked point $AB$. This is a special class of diagrams of the form
\begin{equation}
	\begin{tabular}{cc}
	 \includegraphics[scale=.8]{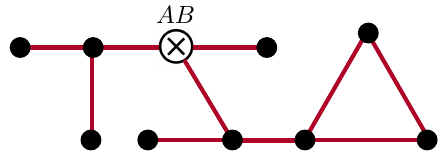} \label{gengr}
	\end{tabular} 
\end{equation}
It is clear that the canonical form for this geometry factorizes into a product of tree branches and the one-cycle core with a branch which ends with a marked point. In other words, we can amputate all tree branches which end on the marked point. 
\begin{equation}
	\begin{tabular}{cc}
	 \includegraphics[scale=.8]{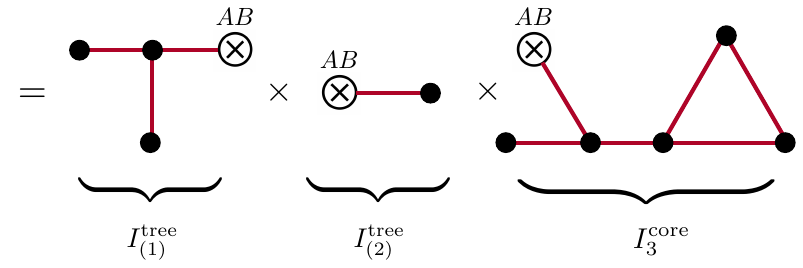}
	\end{tabular} 
\end{equation}
After integration of a general graph we get a product of `tree functions' and the non-trivial function with the cycle,
\begin{equation}
    I(z) = I_{3}^{\rm cycle}(z) \times \prod_k I^{\rm tree}_{(k)}
\end{equation}
As an example, we first look the simplest graph of this kind,
\begin{equation}
	\begin{tabular}{cc}
	 \includegraphics[scale=.77]{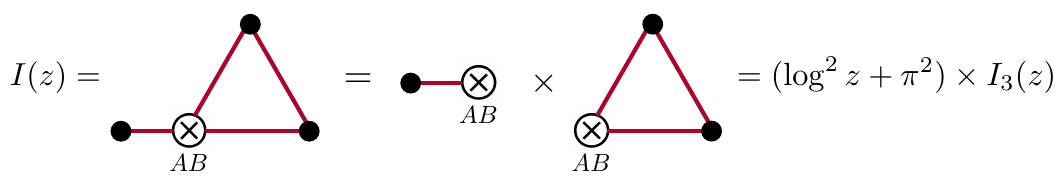}
	\end{tabular} 
\end{equation}
where $I_3(z)$ is given by (\ref{I3z}). Hence we can directly concentrate on the non-trivial piece and consider only graphs where the marked point is at the end of a tree branch (cutting off all trivial tree branches),
\begin{equation}
	\begin{tabular}{cc}
	 \includegraphics[scale=.8]{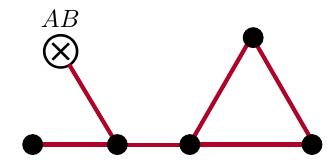}
	\end{tabular} 
\end{equation}
If the marked point is not a part of the triangle cycle, then we can apply a differential equation and get a simpler graph, in complete analogy with tree graphs. At the same time, the boundary condition: vanishing for $z=-1$ still applies. Let us look at the simplest graph of this kind,
\begin{equation}
	\begin{tabular}{cc}
	 \includegraphics[scale=.8]{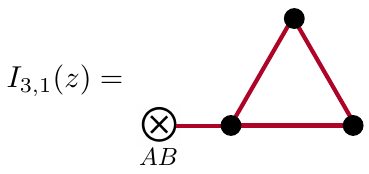}
	\end{tabular} 
\end{equation}
Applying the Laplacian on the frozen loop we solve the same Laplace equation
\begin{equation}
	\begin{tabular}{cc}
	 \includegraphics[scale=.8]{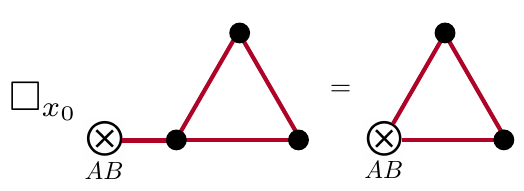}
	\end{tabular} 
\end{equation}
with the boundary condition $I_{3,1}(z)=0$ at $z=-1$ This can be done conveniently in the HPL representation using properties of harmonic polylogarithms \cite{REMIDDI_2000}. More specifically, we can use the fact that $z \partial_z H_{0,n}(z) = H_n(z)$ where $n=\{a_1,a_2,a_3,...,a_n \}$ and so $(z \partial_z)^2 H_{0,0,n}(z) = H_{n}(z)$. Since every term is either of the form $H_{n}(z)$ or a constant with non-zero transcendality, we easily invert every term with respect to this operator, up to any terms destroyed by the operator. And for constants such as $\frac{13\pi^4}{45}$, we use the fact that $\frac{1}{2}(z \partial_z)^2 \log^2(z) = 1$ to add $\pi^4 \log^2(z)$ type terms. As we add more branches, we get $\log^n(z)$ terms, for which we invert the Laplacian using $(z \partial_z)^2 \log^n(z) = n(n-1)\log^{n-2}(z)$. Finally, we have terms that vanish under the Laplacian, which we fix using the boundary condition that the tree numerator must vanish under $z \rightarrow -1$. These vanishing terms are of the form $ C \log(z)+D$, where $C$ and $D$ are constants, and the result reads
\begin{align}
    I_{3,1}(z) &= 8H_{0,0,0,0,0,0}(z)+8H_{0,0,{-}1,0,0,0}(z){-}16H_{0,0,{-}1,{-}1,0,0}(z)+8H_{0,0,{-}2,0,0}(z)- \nonumber\\
    & 8 \zeta_3 (2H_{0,0,{-}1}(z)-H_{0,0,0}(z)) +4 \pi^2(H_{0,0,{-}1,0}(z)-H_{0,0,{-}1,{-}1}(z)+H_{0,0,{-}2}(z))+ \nonumber\\
    &\frac{13 \pi^4}{90} \log^2(z) + C_{3,1} \log(z)+D_{3,1} \,,
\end{align}
To fix the constants that come from the vanishing terms, we impose the boundary condition $I_{3,1}(z{\rightarrow}-1) = 0$ and get
\begin{align}
        C_{3,1} &=-16\zeta_{3,2}+12  \pi ^2 \zeta_3-80  \zeta_5\\
        D_{3,1} & =16 \zeta_{3,3}-16 \zeta^2_3+\frac{289 \pi ^6}{1890}.
\end{align} 
Next, we consider a graph with a longer branch,
\begin{equation}
	\begin{tabular}{cc}
	 \includegraphics[scale=.78]{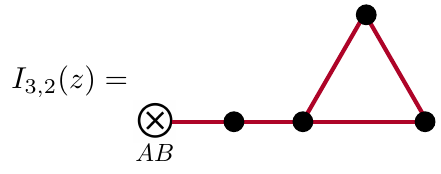}
	\end{tabular} 
\end{equation}
and apply the same Laplacian differential operator,
\begin{equation}
	\begin{tabular}{cc}
	 \includegraphics[scale=.78]{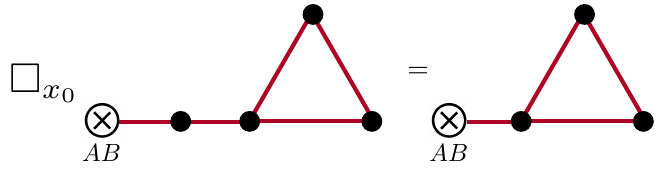}
	\end{tabular} 
\end{equation}
Repeating the procedure, we get
\begin{align}
    I_{3,2}(z) = &8H_{0,0,0,0,0,0,0,0}(z)+8H_{0,0,0,0,{-}1,0,0,0}(z){-}16H_{0,0,0,0,{-}1,{-}1,0,0}(z)+8H_{0,0,0,0,{-}2,0,0}(z) \nonumber\\
    &- 8 \zeta_3 (2H_{0,0,0,0,{-}1}(z)-H_{0,0,0,0,0}(z)) +4 \pi^2(H_{0,0,0,0,{-}1,0}(z)-H_{0,0,0,0,{-}1,{-}1}(z) \nonumber\\
    &{+}H_{0,0,0,0,{-}2}(z)) {+}\frac{13 \pi^4}{1080} \log^4(z) {+} \frac{C_{3,1}}{6} \log^3(z){+} \frac{D_{3,1}}{2}\log(z)^2{+}C_{3,2} \log(z) {+} D_{3,2}\,,
\end{align}
where we kept $C_{3,1}$ and $D_{3,1}$ in that form to make the expression more compact. Evaluating $I_{3,2}(z {\rightarrow} -1)$ and setting it to 0, we can solve for $C_{3,2}$ and $D_{3,2}$ and get
\begin{align}
       C_{3,2} &=-\frac{8}{3}\pi ^2 \zeta_{3,2}-16 \zeta_{5,2}+\frac{127 \pi ^4 \zeta_3}{45}+\frac{8 \pi ^2 \zeta_5}{3}-168 \zeta_7,\\
        D_{3,2} &=\frac{6119 \pi ^8}{75600}+8 \pi ^2 \zeta_{3,3}+16 \zeta_{5,3}+80 \zeta_{6,2}-6 \pi ^2 \zeta^2_3-256 \zeta_5 \zeta_3.
\end{align}
We can repeat this procedure for any size of added branches of this form. Alternatively, we can express the result in terms of ${\rm Li}_n(\frac{z}{1+z})$ or ${\rm Li}_n(\frac{1}{1+z})$ or equivalently in terms of the harmonic polylogarithms $H_n(\frac{z}{1+z})$ and $H_n(\frac{1}{1+z})$. We can invert the Laplacian on such terms using
\begin{equation}
    (z\partial_z)^2 \left( H_{0,0,n}(\frac{z}{1+z})+H_{0,1,n}(\frac{z}{1+z})+H_{1,0,n}(\frac{z}{1+z}) + H_{1,1,n}(\frac{z}{1+z})\right) =H_n(\frac{z}{1+z})\,, 
\end{equation}
and same for $\frac{z}{1+z} \rightarrow \frac{1}{1+z}$. Since integrating the 3-loop triangle with respect to the Laplacian gives functions of the same form, namely $H_{a,n}$, we can keep iterating this procedure and adding more propagators. 

This procedure allows us to solve for any graph of this type. Going beyond that, we would have to evaluate graphs with two branches attached to the triangle, one of them has the marked point, the other one does not. The simplest graph of this kind is 
\begin{center}
	\begin{tabular}{cc}
	 \includegraphics[scale=.8]{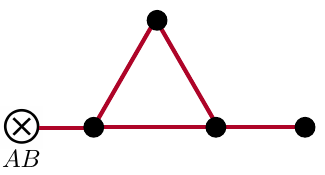}
	\end{tabular} 
\end{center}
which would render a new result which can not be obtained from the core triangle by using the Laplacian operator. The next level of generalization is to consider graphs with arbitrary configurations of three tree branches attached to the triangle, and then graphs with quadrilateral, pentagon, hexagon, etc. cores. We leave this investigation for future work. It is also clear that new ideas are needed to resum all integrated one-cycle graphs (or even some consistent approximation of them), here we just exploited a known result for the three-loop triangle. 

\section{Conclusion and Outlook}
\label{sec:outlook}

In this paper we studied the Amplituhedron picture for four-point scattering amplitudes at all loops. We used the negative geometry expansion of \cite{Arkani-Hamed:2021iya} to determine the canonical form for the all-loop integrand in the (subleading) one-cycle approximation in the context of the loops of loops expansion. We determine the canonical forms by demanding that the numerator vanishes on illegal cuts, producing a unique solution. In the process of determining the form we came up with an interesting organizational principle for the canonical form by further expanding the numerator in terms of certain irreducible building blocks. We believe that the same procedure can be used for the canonical forms of negative geometries with more internal cycles, and perhaps more generally for other positive geometries too. 

There are two main open questions motivated by our work: (1) calculation of higher-cycle canonical forms, (2) determination of the contribution of one-cycle geometries to ${\cal F}(g,z)$, behavior at strong coupling and the gamma cusp. The main tool to tackle (2) will probably be the differential equation method used in \cite{Arkani-Hamed:2021iya} to solve the tree-graph problem, but here we will need to use a different differential operator. If this mission is accomplished, it would open many directions of research and we could start to see a new emergent geometric picture for strong coupling amplitudes.

The related question is how well the loops of loops expansion approximates the full result ${\cal F}(g,z)$. The main reason for dividing the collection of all negative geometries for $\widetilde{\Omega}_L$ into the sectors with fixed number of cycles is obvious: more cycles = more complicated forms, and we wanted to focus on sectors for which we can actually find the solution. But if we really believe this is a good expansion, we might try to introduce a deformation parameter $\xi$ and weigh different $\ell$-cycle sectors with powers $\xi^\ell$,
\begin{equation}
    {\cal F}(g,z,\xi) = {\cal F}^{(0)}(g,z) + \xi {\cal F}^{(1)}(g,z) + \xi^2 {\cal F}^{(2)}(g,z) + \xi^3 {\cal F}^{(3)}(g,z) \dots
\end{equation}
where ${\cal F}^{(k)}(g,z)$ is the $k$-cycle contribution. The physical case is $\xi=1$ where all geometries are weighted equally, while our expansion is around $\xi=0$. Hence a priori it is not clear why this should be a good expansion, but the results of \cite{Arkani-Hamed:2021iya} for the $k=0$ case are optimistic. It would be interesting to see if the purely geometric parameter $\xi$ has some physics analogue as some deformation of the Wilson loop picture. 

\section*{Acknowledgments}

We thank Nima Arkani-Hamed, Dmitry Chicherin, Lance Dixon, Song He, Johannes Henn and Bernd Sturmfels for useful and inspiring discussions. This research received funding from the DOE grant No. SC0009999 and the funds of the University of California.

\appendix

\section{Proof of One-Cycle Result}
\label{sec:proof}

In our construction of the one-cycle integrand, we start with an ansatz that is the sum of a product of $N^{+(-)}$ on the corresponding positive (negative) links $F$ and a remainder $R$. The latter is built out of sums of products of the following objects:
\begin{equation}
    N^a_{i},\ N^b{i},\ N^a_{ij},\ N^b_{ij},\ N^c_{ij}, N^a_{ijkl},\ N^b_{ijkl},\ N^c{ijkl},\ N^a_{ijklmn},\ \cdots 
    \label{eq:allNs}
\end{equation}
We then require that this ansatz for $R$ vanishes on all illegal 2-loop cuts, which disallows certain monomials. Finally, we place the constraint that the whole numerator (fixed links $F$ + remiander $R$) must vanish on the $L$-loop double pole cut. This last condition uniquely fixes all free coefficients and we claim the result \eqref{eq:final1Loop} to be the full one-cycle integrand.

In principle, if a general ansatz is uniquely fixed by requiring it vanishes on all illegal cuts, then the resulting integrand is correct. That being said, constructing a remainder out of the factors \eqref{eq:allNs} is not the most general ansatz. Indeed it is not even the most general ansatz that satisfies all 2-loop conditions. As a result, in order to be sure that our result \eqref{eq:final1Loop} is correct, we need to make a comprehensive list of illegal cuts and then prove that the our result indeed vanishes on all of these cuts. 

Let us begin with a classification of all illegal cuts. There are two types:
\begin{enumerate}
    \item Since the same (up to sign) remainder $R$ works for any set of positive and negative links, the remainder must vanish on any cut that forces a link to have definite sign.
    \item Locality requires that the integrand has no double poles. Any cut that exposes such a double pole (by setting two factors in the denominator equal) is also illegal.
\end{enumerate}

The first type of cut is easy to classify. We simply express the links in some parametrization e.g. \eqref{eq:param}. In these variables, $x_i,y_i,w_i,z_i>0$ in order to be within the 1-loop amplituhedron. As a result we can rewrite a general link suggestively:
\begin{equation}
    \ang{AB_i AB_j}=  \left(w_2 x_1 + w_1 x_2 + y_2 z_1 + y_1 z_2\right) -\left(w_1 x_1+ w_2 x_2 + y_1 z_1 + y_2 z_2\right)\,.
\end{equation}
It is now clear that the list of cuts that force $\ang{AB_i AB_j}$ to be negative (positive) set the first (second) set of terms to zero. We refer to these as 
\begin{enumerate}
    \item $\mathbf{P1}$(i): This is a condition on $x_i, y_i, w_i$ and $z_i$ alone that leads to $\ang{AB_i AB_j}<0$, e.g. $x_i=y_i=z_i=w_i=0$ i.e. the cut where $\ang{AB_i k k{+}1}=0$.
    \item $\mathbf{P2}$(i,j): This is a condition on both variables that sets to $\ang{AB_i AB_j}<0$, e.g. $y_i=w_i=y_j=w_j=0$ i.e. the cut where $\ang{AB_i 12}=\ang{AB_i23}=\ang{AB_j12}=\ang{AB_j23}=0$.
    \item $\mathbf{N1}$(i): This is a condition on $x_i, y_i, w_i$ and $z_i$ alone that leads to $\ang{AB_i AB_j}>0$, e.g. $x_i=y_i=0$ while $w_i, z_i\to\infty$ i.e. the cut where $\ang{AB_i12}=\ang{AB_i14}=0$.
    \item $\mathbf{N2}$(i,j): This is a condition on both sets of variables that sets to $\ang{AB_i AB_j}>0$, e.g. $x_i,y_i,z_j,w_j\to\infty$ i.e. the cut where $\ang{AB_i 23}=\ang{AB_i34}=\ang{AB_j12}=\ang{AB_j14}=0$.
\end{enumerate}

Importantly, the building blocks \eqref{eq:allNs} vanish on many of the cuts:
\begin{enumerate}
    \item $N^{a(b)}_i$ vanishes on $\mathbf{P1}$(i), $\mathbf{N1}$(i) and $\mathbf{N2}$(i,j).
    \item $N^{a(b)}_i N^{a(b)}_j$ vanishes on $\mathbf{P1}$(i), $\mathbf{N1}$(i), $\mathbf{P1}$(j), $\mathbf{N1}$(j), $\mathbf{P2}$(i,j) and $\mathbf{N2}$(i,j).
    \item $N^{a(b)}_{ij\cdots}$ vanishes on $\mathbf{P1}$(i), $\mathbf{P1}$(j) and $\mathbf{P2}$(i,j).
    \item $N^{c}_{ij\cdots}$ vanishes on $\mathbf{N1}$(i), $\mathbf{N1}$(j) and $\mathbf{N2}$(i,j).
\end{enumerate}

Clearly the final result for the remainder \eqref{eq:R1Loop}, only contains monomials in $N$ that vanish on all illegal cuts up to 2-loop order:
\begin{align}
    R^{\rm 1-loop}_L = \bigg\{(2^L{-}4)N_{1}^{(a)}N_{2}^{(a)} \dots N_{L}^{(a)} {-} \sum_{p\in {\rm even}}\sum_{i} \big\{N_{i_1i_2{\dots}i_p}^{(a)}N_{i_1i_2{\dots}i_p}^{(c)}N_{i_p{+}1}^{(a)}\dots N_{i_L}^{(a)}\big\}\bigg\} + (a{\rightarrow} b)\nonumber
\end{align}

The second type of cut is a bit more involved. At 3-loop, we can see what types of cuts are illegal by starting with the zeros of a two links, without loss of generality we choose $\ang{ABCD}$ and $\ang{CDEF}$. We then evaluate $\ang{ABEF}$ on these zeros to find cuts on which setting $\ang{ABCD}$ and $\ang{CDEF}$ to zero accidentally also sets the third propagator to zero i.e. a double pole. In the parametrization \eqref{eq:param}, it is clear that there are only four types of such zeros: $x_i=x, y_i=y, z_i=z$ or $w_i=w$. Interestingly, this fact continues to all $L$-loop, giving us the four types of double pole cuts $\mathbf{DPL}$ that can now be classified into four types: $\ang{AB_i12}=0,\ang{AB_i23}=0,\ang{AB_i34}=0$ or $\ang{AB_i14}=0$.

To see that the $L$-loop result \eqref{eq:R1Loop} vanishes on a $\mathbf{DPL}$ cut, we note that each of the numerators simplifies significantly on these cuts. For example, in the parametrization \eqref{eq:param}, let us consider the building blocks $N$ on the double pole cut where $\ang{AB_i14}=0$, i.e. $x_i=0, y_i=y$. We discuss here this particular cut, but the others follow similarly.
\begin{align}
    &N^+_{ij} =N^-_{ij} = y(z_i+z_j)\,,\qquad N_i^b= N_{i_1i_2\cdots i_{2n}}^{b} =0\,, \nonumber\\
    &N_i^{a} = N_i^c = y z_i\,,\qquad N_{i_1i_2\cdots i_{2n}}^{a}=N_{i_1i_2\cdots i_{2n}}^{c} =y^n\left(\prod_{j\in{\rm even}} z_{i_j}+\prod_{j\in{\rm odd}} z_{i_j} \right)\,.
\end{align}
On this cut, the full numerator \eqref{eq:result1Loop} simplifies:
\begin{align}
    y^L \prod_{i=1}^{L} (z_i+z_{i+1})+y^L (2^L-4) \prod_{i=1}^{L} z_i - y^L \sum_{p\in{\rm even}}\sum_i \left(\prod_{\underset{j<p}{j\in{\rm even}}} z_{i_j}+\prod_{\underset{j<p}{j\in{\rm odd}}} z_{i_j} \right)^2\prod_{j>p} z_{i_j}\,.
\end{align}
We can now consider the coefficient of the two different types of $z_i$ monomials:
\begin{align}
    \prod_{i=1}^{L} z_i &:\ 2+2^L-4 - \sum_{p\in{\rm even}}\sum_i 2 = 2+2^L-4 - \sum_{p\in{\rm even}}2 \frac{L!}{(L-p)!p!}\nonumber\\
    &\ = 2+2^L -4 - 2^L+2 = 0\,,\nonumber\\
    \prod_{\{i_1,\cdots,i_{\frac p2}\}} z_{i}^2 \prod_{\{i_{p{+}1},\cdots,i_{L}\}} z_{i} &:\ 1-1=0\,.
\end{align}
This concludes the proof that our result vanishes on \emph{all} illegal cuts and thus the full numerator \eqref{eq:final1Loop} is indeed the correct one-cycle integrand for planar $\mathcal{N}=4$ SYM at $L$-loop.

\section{Two-loop invariant basis}
\label{sec:2cycle}

Let us determine the two-loop invariant  $R_{1234,13}^{\rm 2-loop}$ of the graph,
\begin{equation}
	\begin{tabular}{cc}
	 \includegraphics[scale=.84]{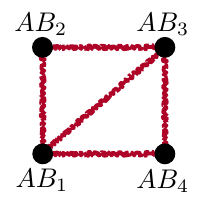}
	\end{tabular} 
\end{equation}
It is useful to organize the basis of terms for the ansatz into three groups
\begin{equation}
{\cal B} = \big\{ {\cal B}_a, {\cal B}_b, {\cal B}_{ab} \big\}\,,
\end{equation}
where ${\cal B}_a$ only contains $N^{(a)}$ and $N^{(c)}$ terms, ${\cal B}_b$ only contains $N^{(b)}$ and $N^{(c)}$ terms and ${\cal B}_{ab}$ contains terms with $a{-}b$ mixing, ie. both $N^{(a)}$ and $N^{(b)}$ terms. Obviously ${\cal B}_b$ can be obtained from ${\cal B}_a$ by relabeling $a\rightarrow b$. 

In the construction of the basis, we follow the following logic: we start with terms that include only one-line invariants $N^{(a,b)}_k$, then also two-line invariants etc. The first term to consider is
\begin{equation}
N^{(a)}_1N^{(a)}_2N^{(a)}_3N^{(a)}_4\,.
\end{equation}
This already satisfies all singlet and also doublet conditions for any pair of lines (even the link $\la AB_2 AB_4\ra$ is covered which we do not need). Based on the little group weights of the numerator, the only term left we can add is $N^{(a,b,c)}_{13}$, hence we add to our basis sets,
\begin{align}
&{\cal B}_a^{(1)} = N^{(a)}_1N^{(a)}_2N^{(a)}_3N^{(a)}_4N_{13}^{(a)}, \qquad
{\cal B}_a^{(2)} = N^{(a)}_1N^{(a)}_2N^{(a)}_3N^{(a)}_4N_{13}^{(c)}\,,\\
&\hspace{3cm} {\cal B}_{ab}^{(1)} = N^{(a)}_1N^{(a)}_2N^{(a)}_3N^{(a)}_4N_{13}^{(b)}\,.
\end{align}
We can also have a mix of $N_j^{(a)}$ and $N_k^{(b)}$ terms. The inequivalent possibilities have `skeletons' built from one-line invariants and then a completion using two-line invariants to satisfy all singlet and double conditions.
\begin{enumerate}
    \item $N_1^{(a)}N_2^{(a)}N_3^{(a)}N_4^{(b)}\qquad \mbox{one option for completion:} \quad N_{13}^{(b)}$
    \item $N_1^{(a)}N_2^{(a)}N_3^{(b)}N_4^{(a)}\qquad \mbox{one option for completion:} \quad N_{13}^{(a)}$
    \item $N_1^{(a)}N_2^{(b)}N_3^{(a)}N_4^{(b)}\qquad \mbox{one option for completion:} \quad N_{13}^{(b)}$
    \item $N_1^{(a)}N_2^{(a)}N_3^{(b)}N_4^{(b)}\qquad \mbox{ no option for completion}$
    \item $N_1^{(b)}N_2^{(a)}N_3^{(b)}N_4^{(a)}\qquad \mbox{one option for completion:} \quad N_{13}^{(a)}$
\end{enumerate}
As a result, we get four new terms that belong to ${\cal B}_{ab}$,
\begin{align}
&{\cal B}_{ab}^{(2)} = N_1^{(a)}N_2^{(a)}N_3^{(a)}N_4^{(b)}N_{13}^{(b)},\qquad 
{\cal B}_{ab}^{(3)} = N_1^{(a)}N_2^{(a)}N_3^{(b)}N_4^{(a)}N_{13}^{(a)}\\
&{\cal B}_{ab}^{(4)} = N_1^{(a)}N_2^{(b)}N_3^{(a)}N_4^{(b)}N_{13}^{(b)},\qquad 
{\cal B}_{ab}^{(5)} = N_1^{(b)}N_2^{(a)}N_3^{(b)}N_4^{(a)}N_{13}^{(a)}\,.
\end{align}
The next case is when we have only three one-line invariants. There are two skeletons which involve only $N_j^{(a)}$,
\begin{enumerate}
    \item $N_1^{(a)}N_2^{(a)}N_3^{(a)}\qquad \mbox{one option for completion:} \quad N_{14}^{(a)}N_{34}^{(c)}$
    \item $N_1^{(a)}N_2^{(a)}N_4^{(a)}\qquad \mbox{no option for completion}$
\end{enumerate}
and four skeletons with the mix of $N_j^{(a)}$ and $N_k^{(b)}$ but none of them can be completed to a valid term,
\begin{enumerate}
    \item $N_1^{(a)}N_2^{(a)}N_3^{(b)}\qquad \mbox{no option for completion}$
    \item $N_1^{(a)}N_2^{(b)}N_3^{(a)}\qquad \mbox{no option for completion}$
    \item $N_1^{(a)}N_2^{(a)}N_4^{(b)}\qquad \mbox{no option for completion}$
    \item $N_1^{(b)}N_2^{(a)}N_4^{(a)}\qquad \mbox{no option for completion}$
\end{enumerate}
As a result we get only one new term in the $a$-sector,
\begin{equation}
{\cal B}_a^{(3)} = N_1^{(a)}N_2^{(a)}N_3^{(a)}N_{14}^{(a)}N_{34}^{(c)}\,.
\end{equation}
Next, we only consider the presence of two one-line invariants, there are three skeletons with only $N_j^{(a)}$,
\begin{enumerate}
    \item $N_1^{(a)}N_2^{(a)}\hspace{0.4cm} \mbox{three options for completion:} \,N_{34}^{(a)}N_{34}^{(c)}N_{13}^{(a,b,c)}$
    \item $N_1^{(a)}N_3^{(a)}\hspace{0.4cm} \mbox{five options for completion:} \,N_{24}^{(a)}N_{24}^{(c)}N_{13}^{(a,b,c)},\, N_{12}^{(a)}N_{24}^{(c)}N_{34}^{(a)},\, N_{12}^{(c)}N_{24}^{(a)}N_{34}^{(c)}$
    \item $N_2^{(a)}N_4^{(a)}\hspace{0.4cm} \mbox{three options for completion:}\,N_{13}^{(a)}N_{13}^{(c)}N_{13}^{(a,b,c)}$
\end{enumerate}
There are two extra terms for the $N_1^{(a)}N_3^{(a)}$ skeleton which include a four-line invariant, namely we can have 
\begin{equation}
    N_1^{(a)}N_3^{(a)} \quad \mbox{completed by}\,\,\, N_{1234}^{(a)}N_{24}^{(c)} \,\,\mbox{or}\,\,N_{1234}^{(c)}N_{24}^{(a)}
\end{equation}
In the mixed sector we also have three skeletons,
\begin{enumerate}
    \item $N_1^{(a)}N_2^{(b)}\qquad \mbox{one option for completion:} \quad N_{34}^{(a)}N_{34}^{(c)}N_{13}^{(b)}$
    \item $N_1^{(a)}N_3^{(b)}\qquad \mbox{one option for completion:} \quad N_{24}^{(a)}N_{24}^{(c)}N_{13}^{(b)}$ 
    \item $N_2^{(a)}N_4^{(b)}\qquad \mbox{one option for completion:} \quad  N_{13}^{(a)}N_{13}^{(c)}N_{13}^{(b)}$
\end{enumerate}
As a result we get new basis terms, in particular in the $a$-sector
\begin{align}
&{\cal B}_a^{(4)} = N_1^{(a)}N_2^{(a)}N_{34}^{(a)}N_{34}^{(c)}N_{13}^{(a)},\qquad 
{\cal B}_a^{(5)} = N_1^{(a)}N_2^{(a)}N_{34}^{(a)}N_{34}^{(c)}N_{13}^{(c)},\\
&{\cal B}_a^{(6)} = N_1^{(a)}N_3^{(a)}N_{24}^{(a)}N_{24}^{(c)}N_{13}^{(a)},\qquad 
{\cal B}_a^{(7)} = N_1^{(a)}N_3^{(a)}N_{24}^{(a)}N_{24}^{(c)}N_{13}^{(c)},\\
&{\cal B}_a^{(8)} = N_1^{(a)}N_3^{(a)}N_{12}^{(a)}N_{24}^{(c)}N_{34}^{(a)},\qquad
{\cal B}_a^{(9)} = N_1^{(a)}N_3^{(a)}N_{12}^{(c)}N_{24}^{(a)}N_{34}^{(c)},\\
&{\cal B}_a^{(10)} = N_2^{(a)}N_4^{(a)}N_{13}^{(a)}N_{13}^{(c)}N_{13}^{(a)},\qquad
{\cal B}_a^{(11)} = N_2^{(a)}N_4^{(a)}N_{13}^{(a)}N_{13}^{(c)}N_{13}^{(c)},\\
&{\cal B}_a^{(12)} = N_1^{(a)}N_3^{(a)}N_{1234}^{(a)}N_{24}^{(c)},\qquad\qquad
{\cal B}_a^{(13)} = N_1^{(a)}N_3^{(a)}N_{1234}^{(c)}N_{24}^{(a)}.
\end{align}
and in the mixed $a-b$ sector,
\begin{align}
&{\cal B}_{ab}^{(6)} = N_1^{(a)}N_3^{(a)}N_{24}^{(a)}N_{24}^{(c)}N_{13}^{(b)},\qquad
{\cal B}_{ab}^{(7)} = N_1^{(a)}N_2^{(a)}N_{34}^{(a)}N_{34}^{(c)}N_{13}^{(b)},\\
&{\cal B}_{ab}^{(8)} = N_2^{(a)}N_4^{(a)}N_{13}^{(a)}N_{13}^{(c)}N_{13}^{(b)},\qquad
 {\cal B}_{ab}^{(9)} = N_1^{(a)}N_2^{(b)}N_{34}^{(a)}N_{34}^{(c)}N_{13}^{(b)},\\
&{\cal B}_{ab}^{(10)} = N_1^{(a)}N_3^{(b)}N_{24}^{(c)}N_{12}^{(b)}N_{34}^{(a)},\qquad {\cal B}_{ab}^{(11)} = N_2^{(a)}N_4^{(b)}N_{13}^{(a)}N_{13}^{(c)}N_{13}^{(b)},\\
&\hspace{3cm}{\cal B}_{ab}^{(12)} = N_1^{(a)}N_3^{(a)}N_{1234}^{(b)}N_{24}^{(c)}\,.
\end{align}
There are only two possible skeleton terms with one one-line invariant each -- either $N_1^{(a)}$ or $N_2^{(a)}$, but none of them can be completed into a valid term with two-line invariants $N_{ij}^{(a,b,c)}$. For $N_1^{(a)}$ we can also use four-line invariants, and we get two more terms in the basis,
\begin{align}
{\cal B}_a^{(14)} = N_1^{(a)}N_{1234}^{(a)}N_{23}^{(c)}N_{34}^{(c)}\,,\qquad
{\cal B}_a^{(15)} = N_1^{(a)}N_{1234}^{(c)}N_{23}^{(a)}N_{34}^{(a)}\,.
\end{align}
There are no terms with no one-line invariants which would match all constraints. The best candidate would be $N_{1234}^{(a)}N_{1234}^{(c)}$ but we are left with only one more term to add to take care of both positive and negative conditions for the $\la AB_1AB_3\ra$ link, and that is not possible. Our final ansatz is then
\begin{equation}
R_{1234,13}^{\rm 2-loop} = \sum_{j=1}^{15} c_j\left({\cal B}_a^{(j)} + {\cal B}_b^{(j)}\right) + \sum_{k=1}^{12} d_k\,{\cal B}_{ab}^{(k)}\,.
\end{equation}
Note that the basis elements are symmetrized according to the graph. For example, two of the properly symmetrized terms are
\begin{align}
    {\cal B}_a^{(3)} &= N_1^{(a)}N_2^{(a)}N_3^{(a)}\left(N_{14}^{(a)}N_{34}^{(c)} + N_{14}^{(c)}N_{34}^{(a)}\right)\,,\\
    {\cal B}_{ab}^{(8)} & = N_1^{(a)} \left(N_2^{(b)}N_{34}^{(a)}N_{34}^{(c)}N_{13}^{(b)} + N_4^{(b)}N_{23}^{(a)}N_{23}^{(c)}N_{13}^{(b)}\right)\nonumber\\ &\hspace{0.5cm} + N_3^{(a)} \left(N_2^{(b)}N_{14}^{(a)}N_{14}^{(c)}N_{13}^{(b)} + N_4^{(b)}N_{12}^{(a)}N_{12}^{(c)}N_{13}^{(b)}\right)\,.
\end{align}
and similarly for all other basis terms -- we need to symmetrize in $AB_1{\leftrightarrow}AB_3$, $AB_2{\leftrightarrow}AB_4$.

\subsection*{Fixing on cuts}

Now we impose cancellation of all illegal cuts, and the cancellation of the double pole for the whole $\Omega^{\rm 2-loop}_G$ form and get a unique solution,
\begin{align}
&c_1 = c_2 = -8,\,\,\, c_3 = -2,\,\,\, c_8 = c_9 = c_{12} = c_{13} = 0,\\
&c_4 = c_5 = c_6 = c_7 = c_{10} = c_{11} = c_{14} = c_{15} = 1, \nonumber\\
& d_1 = d_3 = d_4 = d_5 = 0, \,\,\, d_2 = -4,\,\,\,d_6 = d_7 = d_8 = 0, \,\,\, d_9 = d_{10} = d_{11} = 1,\,\,\, d_{12} = -2\,.\nonumber
\end{align}
This gives us $R_{1234,13}^{\rm 2-loop}$ which completes the calculation of the two-cycle negative geometry. Plugging back the coefficients we find that the structure of the result is very suggestive and gives a hope that the closed form exists also for graphs with more cycles.

\bibliographystyle{JHEP}
\bibliography{refs.bib}

\end{document}